\documentclass[%
reprint,
showpacs,preprintnumbers,
 amsmath,amssymb,
 aps,
 pre,
floatfix,
]{revtex4-1}

\usepackage{graphicx}% Include figure files
\usepackage{dcolumn}% Align table columns on decimal point
\usepackage{bm}% bold math
\usepackage{upgreek}
\usepackage{color}
\usepackage[normalem]{ulem}
\begin{document}

\preprint{APS/123-QED}

\title{Pushing the Limits of Monte Carlo Simulations for the 3d Ising Model}

\author{Alan M. Ferrenberg$^1$}
\email{alan.ferrenberg@miamioh.edu}
\author{Jiahao Xu$^2$}
\author{David P. Landau$^2$}
\email{dlandau@physast.uga.edu}
\affiliation{$^1$Information Technology Services and Department of Chemical, Paper \rm{\&} \textit{Biomedical Engineering, Miami University, Oxford, OH 45056 USA\\
$^2$Center for Simulational Physics, University of Georgia, Athens, GA 30602 USA}}

\date{\today}

\begin{abstract}
While the 3d Ising model has defied analytic solution, various numerical 
methods like Monte Carlo, MCRG and series expansion have provided precise 
information about the phase transition. Using Monte Carlo simulation that employs the Wolff cluster flipping algorithm with both 32-bit and 53-bit random number generators and data analysis with histogram reweighting and 
quadruple precision arithmetic,
we have investigated the critical behavior of the simple cubic Ising Model, with 
lattice sizes ranging from $16^3$ to $1024^3$. By analyzing data with 
cross correlations between various thermodynamic quantities obtained from 
the same data pool, e.g. logarithmic derivatives of magnetization and 
derivatives of magnetization cumulants, we have obtained the critical 
inverse temperature $K_c=0.221\,654\,626(5)$ and the critical exponent of 
the correlation length $\nu=0.629\,912(86)$ with precision that exceeds all previous Monte Carlo estimates. 

%\begin{description}
%\end{description}

\end{abstract}



\maketitle

%\tableofcontents

%%%%%%%%%%%%%%%%%%%%%%%%%%%%%%%%%%%%%%%%%%%%%%%%%%%%%%%%%%%%%%%%%%%%%%%%%%
% section 1: introduction
\section{\label{sec:level1}introduction}
The Ising Model~\cite{Ising}
 has played a seminal role in the theory of phase transitions, 
and has served as a testing ground for innumerable numerical and theoretical approaches. 
Although it has been solved in one- and two-dimension~\cite{Ising,Onsager}, its 
analytic solution for three-dimension is still a mystery. Nevertheless, by the
end of the last century various numerical methods like 
Monte Carlo~\cite{Ferrenberg1991}, nonequilibrium relaxation Monte Carlo~\cite{Ito},
Monte Carlo renormalization group~\cite{blote1996,Pawley_MCRG}, 
field theoretic methods~\cite{expan1998,expan2008} 
and high-temperature series expansions~\cite{HT_expan2002} 
have provided precise information about the nature of the 
phase transition~\cite{CB_review} and critical exponents, although in some cases the results did not agree within the error bars.  In addition, Rosengren made an ``exact conjecture'' for the critical temperature for the 3d Ising model~\cite{Rosengren_exact} and the precision of numerical calculations was insufficient to determine if this prediction was correct. Fisher, however, pointed out that a number of other ``exact conjectures'' could be derived that gave quite similar numerical values~\cite{Fisher_1995}. Hence, while rather precise values existed for the 3d Ising critical temperature, there were still unanswered questions.  (For a rather complete review of results prior to 2002 see Ref.~\cite{CB_review}.)

Over the past decade or so, several new developments appeared that
reinvigorated interest in the critical behavior of the 3d Ising model.
Recently, the conformal bootstrap method, using the constraints of
crossing symmetry and unitarity in conformal field theories, has given
unparalleled precision in the estimates for the critical exponent
$\nu$ for the 3d Ising
model~\cite{conformal_BT2012,conformal_BT2014,conformal_BT2016}.  New
Monte Carlo simulation based, in part, on non-perturbative
approaches~\cite{Hasenbusch2010,MC_non_per1,MC_non_per}, and tensor
renormalization group theory with high-order singularity value
decomposition~\cite{tensor_RG_HO} have also yielded very precise
results. In clarifying work, Wu, McCoy, Fisher, Chayes and Perk~\cite{Wu_comment,Wu_rejoinder,Fisher_comment2016} gave very convincing arguments that a supposed ``exact'' solution was simply wrong.

Precise numerical estimates for various critical properties play an
important role as a testing ground for developing theories and
supposed exact solutions, and Monte Carlo simulation is potentially one
of the best suited methods for delivering quantitative information
about the critical behavior.  In this paper, we present the results
of high-precision Monte Carlo simulations  of critical behavior in the 3d
Ising model, using histogram reweighting
techniques~\cite{histogram_Ferrenberg1988,histogram_Ferrenberg1989},
cross correlation
analysis~\cite{cross_correlation2009,cross_correlation2010} and
finite-size scaling methods~\cite{FSS_Fisher1971, FSS_Fisher1972,
  FSS_Barber1983, FSS_Privman} to obtain high resolution estimates for
the critical coupling and critical exponents.

%%%%%%%%%%%%%%%%%%%%%%%%%%%%%%%%%%%%%%%%%%%%%%%%%%%%%%%%%%%%%%%%%%%%%%%%%%
% section 2: model and methods
%\newpage
\section{\label{sec:level1}model and methods}

%%%%%%%%%%%%%%%%%%%%%%%%%%%%%%%%%%%%%%%%%%%%%%%%%%%%%%%%%%%%%%%%%%%%%%%%%%
% subsection: three-dimensional ising model
\subsection{\label{sec:level2}Three-dimensional Ising model}
We have considered the simple cubic, ferromagnetic Ising model with 
nearest-neighbor interactions on $L \times L \times L$ lattices with 
periodic boundary conditions. Each of the lattice sites $i$ has a spin, 
$\sigma_i$, which can take on the values $\sigma_i = +1$ for 
spin up and $\sigma_i = -1$ for spin down. The interaction Hamiltonian 
is given by
\begin{equation}
\mathcal{H} = -J\sum_{\langle i,j \rangle}\sigma_i\sigma_j,
\end{equation}
where $\langle i,j \rangle$ denotes distinct pairs of nearest-neighbor sites and $J$ is 
the interaction constant. We also define the dimensionless energy $E$ as
\begin{equation}
E = - \sum_{\langle i,j \rangle}\sigma_i\sigma_j.
\end{equation}

In discussing the critical properties of the Ising model, it is easier 
to deal with the inverse temperature, so we define the dimensionless 
coupling constant $K = J/k_BT$ and use $K$ for the discussion.

%%%%%%%%%%%%%%%%%%%%%%%%%%%%%%%%%%%%%%%%%%%%%%%%%%%%%%%%%%%%%%%%%%%%%%%%%%
% subsection: Monte Carlo sampling method
\subsection{\label{sec:level2}Monte Carlo sampling method}
We have simulated $L \times L \times L$ simple cubic lattices using the 
Wolff cluster flipping algorithm \cite{Wolff}. Single clusters are grown 
and flipped sequentially. Bonds are drawn to all nearest neighbors of the growing cluster with probability
\begin{equation}
p = 1 - e^{-2K\delta_{\sigma_i\sigma_j}}
\end{equation}

\noindent To accelerate the Wolff algorithm, we calculated the energy and magnetization 
by only looking at the spins that actually get flipped in the process. 
To do that, rather than flipping spins immediately we temporarily set them equal to zero 
and keep a list of those spins. By setting the spins in that cluster equal 
to zero we don't calculate the internal energy of the cluster since the 
energy change only comes from the edges of the 
cluster. The magnetization change, however, is related to the number of spins in the 
cluster. After calculating the changes, we go back and set all of the 
``zeroed'' spins to their correct value (flipped from their original value). 
So we calculate the energy and magnetization once, then add the changes to 
them to get the new values. 

In the simulation, a new random number was generated for each bond update, 
using the Mersenne Twister random number generator \cite{Mersenne_Twister}. 
We have implemented the Mersenne Twister algorithm with using both 32-bit word 
length and 53-bit word length. 

The simulations were performed at $K_0=0.221\,654$, which is an estimate for 
the critical inverse temperature $K_c$ by MCRG analysis \cite{Pawley_MCRG} and also used in an earlier, high resolution Monte Carlo study~\cite{Ferrenberg1991}. 
Data were obtained for lattices with $16 \leq L \leq 768$, after 
$2 \times 10^5$ Wolff steps were discarded for equilibrium. Even for the largest lattice size, $L=1024$, the system had reached the equilibrium value of the energy by $130\,000$ cluster steps, and the simulation was then run another ten times the equilibrium relaxation time before data accumulation began. Actual lattice sizes studied were $L=$ 16, 24, 32, 48, 64, 80, 96, 112, 128, 144, 160, 192, 256, 384, 512, 768, and 1024.  For $L \leq 768$ we started from an ordered state and the relaxation to equilibrium was less than a thousand Wolff flips in all cases~\cite{Ito_Kohring_1993}.  Our procedure insured not only that equilibrium had been reached but that also correlation with the initial state had been lost.  For $L = 1024$ we began with random states but our procedure insured that the system had reached equilibrium and that more than 10 times the equilibrium relaxation time had elapsed before data were taken. We performed between 6000 runs to $12\,000$ runs of $5 \times 10^6$ measurements for each lattice size. In total, 
we have used around $2 \times 10^7$ CPU core hours and generated more than 5TB 
data using 5 different Linux clusters. For the largest lattice ($L = 1024$), the run length for a single 
run is around 4000 times the correlation time for the internal energy, and the 
average cluster size is around $1.1 \times 10^6$.  

%%%%%%%%%%%%%%%%%%%%%%%%%%%%%%%%%%%%%%%%%%%%%%%%%%%%%%%%%%%%%%%%%%%%%%%%%%
% subsection: Histogram reweighting
\subsection{\label{sec:level2}Histogram reweighting}
One limitation on the resolution of Monte Carlo simulations near phase 
transition is that  many runs must be performed at different temperature 
to precisely locate the peaks in response functions. Using histograms we  
can extract more information from Monte Carlo simulations 
\cite{histogram_Ferrenberg1988, histogram_Ferrenberg1989}, because samples 
taken from a known probability distribution can be translated into samples 
from another distribution over the same state space.

An importance sampling Monte Carlo simulation (in our case using cluster flipping as described above) is first carried out at the inverse 
temperature $K_0$ to generate configurations with a probability 
proportional to the Boltzmann weight, $\exp(-K_0E)$. The probability of 
simultaneously observing the system with total (dimensionless) energy $E$ and 
total magnetization $M$ is,
\begin{equation}
P_{K_0} = \frac{1}{Z(K_0)} W(E,M) \exp(K_0E),
\end{equation}
where $Z(K_0)$ is the partition function, and $W(E,M)$ is the number of 
configurations with energy $E$ and magnetization $M$.
Then, a histogram $H_0(E,M)$ of the energy and the magnetization at $K_0$ is constructed to provide an estimate for the equilibrium probability distribution. Thus,
\begin{equation}
H_0(E,M) = \frac{N}{Z(K_0)} \tilde{W}(E,M) \exp(-K_0E),
\end{equation}
where $\tilde{W}(E,M)$ is an estimate for the true density of states $W(E,M)$, $N$ is the number of measurements made. 
In the limit of an infinite-length run, we can replace $W(E,M)$ with 
$\tilde{W}(E,M)$, which will yield the relationship between the histogram 
measured at $K = K_0$ and the (estimated) probability distribution for arbitrary 
$K$,
\begin{equation}
P_K(E,M) = \frac{H_0(E,M) e^{\Delta K E}}{\sum_{E,M} H_0(E,M) e^{\Delta K E}},
\end{equation}
where $\Delta K = K_0 - K$. Based on $P_K(E,M)$, we can calculate the average 
value of any function of $E$ and $M$, $f(E,M)$,
\begin{equation}
\langle f(E,M) \rangle_K = \sum_{E,M} f(E,M) P_K(E,M)
\end{equation}

As $K$ can be varied continuously, the histogram method is able to locate the 
peaks for different thermodynamic derivatives precisely (e.g. using the golden-section search technique \cite{golden_section}), and it provides an 
opportunity to study the critical behavior using Monte Carlo with high resolution.

%%%%%%%%%%%%%%%%%%%%%%%%%%%%%%%%%%%%%%%%%%%%%%%%%%%%%%%%%%%%%%%%%%%%%%%%%%
% subsection: Quantities to be analyzed
\subsection{\label{sec:level2}Quantities to be analyzed}
\label{quantities}
Ferrenberg and Landau~\cite{Ferrenberg1991} showed that the critical exponent $\nu$ 
of the correlation length can be estimated more precisely from Monte Carlo simulation data if multiple quantities, including traditional quantities which still have 
the same critical properties, are included. The logarithmic derivative of 
any power of the magnetization
\begin{equation}
\frac{\partial \ln{\langle|m|^i\rangle}}{\partial K} = 
\frac{1}{\langle|m|^i\rangle} 
\frac{\partial \langle|m|^i\rangle} {\partial K} = 
\frac{\langle|m|^i E \rangle} {\langle|m|^i\rangle} - 
\langle E \rangle,
\end{equation}
for $i = 1,2,...$, can yield an estimate for $\nu$ and we
have considered the logarithmic derivatives of $\langle|m|\rangle$,  
$\langle|m|^2\rangle$, $\langle|m|^3\rangle$ and 
$\langle|m|^4\rangle$ in this analysis. We also included (reduced) magnetization 
cumulants $U_{2i}$ \cite{cumulant_Binder1981} defined by
\begin{equation}
U_{2i} = 1 - \frac{\langle|m|^{2i}\rangle} 
{3\langle| m |^i\rangle^2}, \quad i = 1,2,3,...
\end{equation}
whose derivatives with respect to $K$ can also be used to estimate $\nu$. In this analysis we have considered the second-order, 
fourth-order and sixth-order cumulants $U_2$, $U_4$ and $U_6$.

Once $\nu$ is determined, we can estimate the inverse critical temperature 
$K_c(L)$ from the locations of the peaks in the above quantities.  Apart from those quantities, we can also use the specific heat
\begin{equation}
C = K^2 L^{-d} ( \langle E^2 \rangle - \langle E \rangle^2 ),
\end{equation}
the coupling derivative of $|m|$,
\begin{equation}
\frac {\partial \langle |m| \rangle} {\partial K} 
= \langle |m|E \rangle - \langle |m| \rangle \langle E \rangle,
\end{equation}
the finite-lattice susceptibility,
\begin{equation}
\chi' = K L^d (\langle |m|^2 \rangle - \langle |m| \rangle^2),
\label{chi}
\end{equation}
and the zero of the fourth-order energy cumulant
%\begin{eqnarray}
%Q_4 &=& \langle E^4 \rangle 
%- 3\langle E^2 \rangle^2 
%- 4\langle E \rangle \langle E^3 \rangle \nonumber \\
%&& +~12\langle E \rangle^2 \langle E^2 \rangle
%- 6\langle E \rangle^4
%\end{eqnarray}
\begin{equation}
Q_4 = 1 - \frac{\langle (E - \langle E \rangle)^{4} \rangle} 
{3\langle (E - \langle E \rangle)^{2} \rangle^2} .
\end{equation}
Note that in Eq.~(\ref{chi}), it is the finite-lattice susceptibility, 
not the ``true'' susceptibility calculated from the variance of $m$, 
$\chi = K L^d (\langle m^2 \rangle - \langle m \rangle^2)$.  The 
``true'' susceptibility cannot be used to determine $K_c(L)$ as it has no 
peak for finite systems. For sufficiently long runs, $\langle m \rangle = 0 $ for zero magnetic field ($h = 0$) so 
that any peak in $\chi$ is merely due to the finite statistics of the 
simulation.

We have calculated all of the above quantities by using the GCC Quad-Precision Math 
Library which provides quadruple (128 bit) precision.

%%%%%%%%%%%%%%%%%%%%%%%%%%%%%%%%%%%%%%%%%%%%%%%%%%%%%%%%%%%%%%%%%%%%%%%%%%
% subsection: Finite-size scaling analysis
\subsection{\label{sec:level2}Finite-size scaling analysis}
At a second order phase transition the critical behavior of a system in the 
thermodynamic limit can be extracted from the size dependence of the singular 
part of the free energy density.  This finite size scaling theory was first developed by Fisher
\cite{FSS_Fisher1971, FSS_Fisher1972, FSS_Barber1983, FSS_Privman}.

According to finite-size scaling theory, and assuming homogeneity, hyperscaling and using 
$L$ (linear dimension) and $T$ (temperature) as variables, the free energy of a system 
is described by the scaling ansatz,
\begin{equation}
F(L,T) = L^{-(2-\alpha)/\nu}\mathcal{F}(\varepsilon L^{1/\nu}, hL^{(\gamma+\beta)/\nu}),
\label{freeEnergy}
\end{equation}
where $\varepsilon = (T - T_c)/T_c$ ($T_c$ is the infinite-lattice critical 
temperature) and $h$ is the magnetic field. The critical exponents $\alpha$, 
$\beta$, $\gamma$ and $\nu$ assume their infinite lattice values. The choice 
of the scaling variable $x = \varepsilon L^{1/\nu}$ is motivated by the 
observation that the correlation length, which diverges as $\varepsilon^{-\nu}$ 
as the transition is approached, is limited by the lattice size $L$. The 
various thermodynamic properties can be determined from Eq.~(\ref{freeEnergy}) 
and have corresponding scaling forms, e.g.,
\begin{eqnarray}
&& m = L^{-\beta/\nu}\mathcal{M}^0(\varepsilon L^{1/\nu}), \\
&& \chi = L^{\gamma/\nu} \mathcal{\chi}^0(\varepsilon L^{1/\nu}) + b_{\chi}, \\
&& C = L^{\alpha/\nu} \mathcal{C}^0(\varepsilon L^{1/\nu}) + b_C, \label{specific_heat_scaling}
\end{eqnarray}
where $\mathcal{M}^0(x)$, $\mathcal{\chi}^0(x)$ and $\mathcal{C}^0(x)$ are 
scaling functions, and $b_{\chi}, b_C$ are analytic background terms. Because we are interested in zero-field properties 
($h = 0$), $x$ is the only relevant thermodynamic variable.

A number of different practical implementations based on FSS schemes have 
been derived and successfully applied to the analysis of the critical 
phenomena \cite{Ferrenberg1991, CB_review, Hasenbusch2010}. In our analysis, we 
determine the effective transition temperature very precisely based on the 
location of peaks in multiple thermodynamic quantities as discussed in Sec.~\ref{quantities}.

Take the specific heat $C$ for example, for a finite lattice, the peak 
occurs at the temperature where the scaling function $\mathcal{C}^0$ is 
maximum, i.e., when 
\begin{equation}
\frac {\partial \mathcal{C}^0(x)} {\partial x} \biggr|_{x = x^*} = 0.
\end{equation}
The temperature corresponding to the peak is the finite-lattice (effective) 
transition temperature $T_c(L)$, on the condition $x = x^*$ varies with 
$L$ asymptotically as
\begin{equation}
T_c(L) = T_c + T_c x^* L^{-1/\nu}.
\label{effective}
\end{equation}

The finite-size scaling ansatz is valid only for sufficiently large lattice 
size, $L$, and temperatures sufficiently close to $T_c$. Corrections to scaling and finite-size scaling must be taken into account for smaller systems and temperatures 
away from $T_c$. Basically, there are two kinds of correction terms, one is 
due to the irrelevant scaling fields which can be expressed in terms of an 
exponent $\theta$ leading to additional terms like 
$a_1 \varepsilon^\theta + a_2 \varepsilon^{2\theta} + \cdots$, 
while the other is due to the non-linear scaling fields which can be expressed 
like $b_1 \varepsilon^1 + b_2 \varepsilon^2 + \cdots$. The temperatures that we 
consider in our analysis differ from $T_c$ (or $\varepsilon = 0$) by amounts 
proportional to $L^{-1/\nu}$ (Eq.~(\ref{effective})), so that the correction 
terms can be expressed by the power-law $a_1 L^{-\theta/\nu} + a_2 L^{-2\theta/\nu}$  and $b_1 L^{-1/\nu} + b_2 L^{-2/\nu}$.

If we take correction terms into account, the estimate for $T_c(L)$ can 
be expressed to be
\begin{flalign}
  T_c(L) & = T_c  + A_0' L^{-1/\nu} (1 + A_1' L^{-\omega_1} + A_2' L^{-2\omega_1} + \cdots \nonumber \\
  & + B_1' L^{-\omega_2} + B_2' L^{-2\omega_2} + \cdots + C_1' L^{-(\omega_1 + \omega_2)} + \cdots \nonumber \\
  & + D_1' L^{-\omega_\nu} + D_2' L^{-2\omega_\nu} + \cdots + E_1' L^{-\omega_{NR}} + \cdots)
\end{flalign}
where $\omega_i \; (i = 1,2,...)$ are the correction exponents, $\omega_\nu = 1 / \nu$ is the correction exponent corresponding to the non-linear scaling fields \cite{Fisher_correction}, and $\omega_{NR}$ is the correction exponent due to the rotational invariance of the lattice \cite{Massimo_wnr}. As we have defined the coupling as $K = J/k_BT$, $K_c(L)$ can be expressed as
\begin{flalign}
K_c(L) & = K_c +  A_0 L^{-1/\nu} (1 + A_1 L^{-\omega_1} + A_2 L^{-2\omega_1} + \cdots \nonumber \\
    & + B_1 L^{-\omega_2} + B_2 L^{-2\omega_2} + \cdots + C_1 L^{-(\omega_1 + \omega_2)} + \cdots \nonumber \\
    & + D_1 L^{-\omega_\nu} + D_2 L^{-2\omega_\nu} + \cdots + E_1 L^{-\omega_{NR}} + \cdots)
\label{Kc}
\end{flalign}

Rather than using Eq.~(\ref{Kc}) to estimate $K_c$ directly, we can first 
estimate the critical exponent $\nu$ using the quantities discussed in Sec.~\ref{quantities}.
After obtaining a precise estimate for $\nu$, we can insert it 
into Eq.~(\ref{Kc}), so that there is one less unknown parameter to do the 
non-linear fit to Eq.~(\ref{Kc}).

To estimate $\nu$ precisely, we can use the following critical scaling form 
without the prior knowledge of the transition coupling $K_c$
\begin{flalign}
\frac{\partial U_{2i}}{\partial K}\biggr|_{\max} & = 
 U_{i,0} L^{1/\nu} (1 + a_1 L^{-\omega_1} + a_2 L^{-2\omega_1} + \cdots \nonumber \\
 & + b_1 L^{-\omega_2} + b_2 L^{-2\omega_2} + \cdots + c_1 L^{-(\omega_1 + \omega_2)} + \cdots \nonumber \\
 & + d_1 L^{-\omega_\nu} + d_2 L^{-2 \omega_\nu} + \cdots  + e_1 L^{-\omega_{NR}} + \cdots)
\label{dU}
\end{flalign}

\begin{flalign}
\frac{\partial \ln{\langle|m|^i\rangle}}{\partial K}\biggr|_{\max} = 
D_{i,0} L^{1/\nu} (1 + a_1 L^{-\omega_1} + a_2 L^{-2\omega_1} + \cdots \nonumber \\
 + \; b_1 L^{-\omega_2} + b_2 L^{-2\omega_2} + \cdots + c_1 L^{-(\omega_1 + \omega_2)} + \cdots \nonumber \\
 + \; d_1 L^{-\omega_\nu} + d_2 L^{-2 \omega_\nu} + \cdots  + e_1 L^{-\omega_{NR}} + \cdots) \;
\label{dlnm}
\end{flalign}

\noindent Once $\nu$ is determined from the fit of Eq.~(\ref{dU}) and Eq.~(\ref{dlnm}), 
we can estimate the critical inverse temperature $K_c$ with a fixed value 
of $\nu$

Another method which can be used to determine the inverse transition 
temperature is Binder's 4th order cumulant crossing technique \cite{cumulant_Binder1981}. 
As the lattice size $L \rightarrow \infty$, the fourth-order magnetization 
cumulant $U_4 \rightarrow 0$ for $K < K_c$ and $U_4 \rightarrow 2/3$ for 
$K > K_c$. $U_4$ can be plotted as a function of $K$ for different lattice 
sizes, and the location of the intersections between curves for the two lattice sizes is given by
\begin{flalign}
K_{\text{cross}}(L,b) = K_c 
+& a_1 L^{-1/\nu-\omega_1} \biggr(\frac{b^{-\omega_1}-1}{b^{1/\nu}-1}\biggr) \nonumber\\
+& a_2 L^{-1/\nu-\omega_2} \biggr(\frac{b^{-\omega_2}-1}{b^{1/\nu}-1}\biggr) 
+ \cdots,
\label{eq_Kcross}
\end{flalign}
where $L$ is the size of the smaller lattice, $b = L'/L$ is the ratio of 
two lattice sizes, and $\omega_1$, $\omega_2$ are correction exponents in 
the finite-size scaling formulation. 

%%%%%%%%%%%%%%%%%%%%%%%%%%%%%%%%%%%%%%%%%%%%%%%%%%%%%%%%%%%%%%%%%%%%%%%%%%
% subsection: Jackknife method with cross correlations
\subsection{\label{sec:level2}Jackknife method with cross correlations}
Ideally, in a Monte Carlo simulation, a configuration only depends 
on the previous configuration, but in practice, it is also 
likely to be correlated to earlier configurations. Generally, the farther 
away two configurations are, the less correlation. 
Because measurements in the 
time-series are correlated, the fluctuations appear 
smaller than they should be. To deal with this issue, we can consider blocks 
of the original data, and use jackknife resampling \cite{jackknife}.

An important advance was made by Weigel and Janke~\cite{cross_correlation2009, cross_correlation2010} via the seminal observation that there could be significant cross correlation between different quantities that could lead to systematic bias in the estimates of critical quantities extracted from the data.

Suppose we have a set (sample) of $n$ measurements of a random variable 
${\bf x} = (x_1, x_2, \cdots , x_n)$, and an estimator 
$\hat{\theta} = f({\bf x})$. To estimate the value and error of 
$\hat{\theta}$ the jackknife focuses on the samples that leave out one 
measurement at a time. We define the jackknife average, $x_i^J$ by,
\begin{equation}
x_i^J = \frac {1} {n-1} \sum_{j \neq i} x_j,
\end{equation}
where $i = 1, 2, \cdots , n$, so $x_i^J$ is the average of all the $x$ values 
except $x_i$. Similarly, we define
\begin{equation}
\hat{\theta}_i^J = f(x_i^J).
\end{equation}
The jackknife estimate of 
$\hat{\theta} = f({\bf x})$ is the average of $\hat{\theta}_i^J$, i.e.
\begin{equation}
\bar{\theta} = \frac {1} {n} \sum_{i=1}^n \hat{\theta}_i^J
= \frac {1} {n} \sum_{i=1}^n f(x_i^J),
\label{jackknife_estimate}
\end{equation}
and the jackknife error $\sigma(\hat{\theta})$, is given by, 
\begin{equation}
\sigma(\hat{\theta}) = \biggr[ \frac {n-1} {n} 
\sum_{i=1}^n (\hat{\theta}_i^J - \bar{\theta})^2 \biggr]^{1/2}
\label{jackknife_error}
\end{equation}

\noindent In Eq.~(\ref{jackknife_estimate}) and Eq.~(\ref{jackknife_error}), each 
data block has only one element, but generally there can be multiple 
adjacent elements in each block. For example, we can have $n$ data blocks, 
where each block has $N_b = N / n$ adjacent elements ($N$ is the total 
number of measurements in the time series).

%%%%%%%%%%%%%%%%%%%%%%%%%%%%%%%%%%%%%%%%%%%%%%%%%%%%%%%%%%%%%%%%%%%%%%%%%%
% subsection: Cross correlation
%\subsection{\label{sec:level2}Cross correlation}
When attempting to extract the parameter $\hat{\theta}$ based on multiple 
estimates $\hat{\theta}^{(k)} (k = 1,2,...,m)$ from the same original 
time-series data, Weigel and Janke~\cite{cross_correlation2009, cross_correlation2010} showed that there could be significant cross correlation between 
estimates $\hat{\theta}^{(k)}$ and $\hat{\theta}^{(l)}$. For example, we can 
determine a number of estimates for $\nu$ from Eq.~(\ref{dU}) and 
Eq.~(\ref{dlnm}). Denoting them $\nu^{(k)} (k=1,2,...,m)$, we obtain 
different $\nu^{(k)}$ from different quantities, although they are all calculated from
the same configurations of the system.

To reduce the cross correlation effectively, we considered the jackknife 
covariance matrix ${\bf{G}} \in \mathbb{R}^{m \times m}$~\cite{jackknife}. For a number of estimates 
$\hat{\theta}^{(k)}$, the $r^{\rm{th}}$ row, $c^{\rm{th}}$ column entry of 
matrix ${\bf{G}}$ is given by,
\begin{equation}
{\bf{G}}_{rc} (\hat{\theta}) = \frac {n-1} {n} \sum_{i=1}^n 
(\hat{\theta}_i^{J,(r)} - \bar{\theta}^{(r)}) 
(\hat{\theta}_i^{J,(c)} - \bar{\theta}^{(c)}).
\end{equation}

The $m$ different estimates $\hat{\theta}^{(k)} (k=1,2,...,m)$ for 
the same parameter $\hat{\theta}$, should have the same expectation value. 
So the estimated value for $\hat{\theta}$ can be determined by a linear 
combination, 
\begin{equation}
\bar{\theta} = \sum_{k=1}^m \alpha_k \hat{\theta}^{(k)}. 
\end{equation}
where $\sum_k \alpha_k = 1$. Based on the cross correlation analysis from 
Ref.~\cite{cross_correlation2009, cross_correlation2010}, a Lagrange 
multiplier can be introduced, where the constraint $\sum_k \alpha_k = 1$ is 
enforced, to minimize the variance,
\begin{equation}
\sigma^2(\hat{\theta}) = \sum_{k=1}^m \sum_{l=1}^m \alpha_k \alpha_l 
(\langle \hat{\theta}^{(k)} \hat{\theta}^{(l)} \rangle - 
\langle \hat{\theta}^{(k)} \rangle \langle \hat{\theta}^{(l)} \rangle).
\end{equation}
The optimal choice for the weights is,
\begin{equation}
\alpha_k = \frac {\sum_{l=1}^m [ {\bf{G}}(\hat{\theta})^{-1} ]_{kl}} 
{\sum_{k=1}^m \sum_{l=1}^m [ {\bf{G}}(\hat{\theta})^{-1} ]_{kl}},
\label{eq_weights}
\end{equation}
where ${\bf{G}}(\hat{\theta})^{-1}$ is the inverse of the covariance matrix. 
Traditionally, the weights are bounded to be $0 \leq \alpha_k \leq 1$; 
however, the optimal choices given in Eq.~(\ref{eq_weights}) are the more 
general unbounded weights, which can be negative. The negative weights may 
lead to the average lying outside the range of individual estimates, where 
individual variances are connected due to cross correlations. Thus, they can 
help alleviate the effect of cross correlations. 

Based on the optimal choice for the weights, the variance can be expressed 
by,
\begin{equation}
\sigma^2(\hat{\theta}) = \frac {1} 
{\sum_{k=1}^m \sum_{l=1}^m [ {\bf{G}}(\hat{\theta})^{-1} ]_{kl}}.
\end{equation}

%%%%%%%%%%%%%%%%%%%%%%%%%%%%%%%%%%%%%%%%%%%%%%%%%%%%%%%%%%%%%%%%%%%%%%%%%%
% subsection: Sampling algorithms and testing methodology
\subsection{\label{sec:level2}Testing methodology and quality control}

A new challenge that arises at the level of accuracy of this study is the finite
precision of the pseudorandom number generator and the restriction
this puts on the temperatures that can be simulated.  In the Wolff
algorithm, the probability of adding a spin to the cluster is related
to $K$ by
\[
p = 1 - e^{-2K\delta_{\sigma_i\sigma_j}}
\]
When this probability is converted to a 32-bit unsigned number for
comparison with pseudorandom numbers generated in the simulation it is
truncated from 1537987121.70821 to 1537987121.  If that is reconverted
back into a value of $K$ the result differs from 0.221 654 in the 10th
decimal place.  For the largest system sizes, this is only a factor of
20 smaller than the statistical error.  By performing simulations with
a 53-bit pseudorandom number generator we have verified that this is
not significant for the current analysis, but for future studies of
larger systems and/or higher precision, a 32-bit random number
generator would not be sufficient.  For the data analysis we used the
corrected effective $K_0$ instead of 0.221 654 and for $L=1024$ we used
the multiple-histogram method~\cite{histogram_Ferrenberg1989} to combine results for the 32 and
53-bit pseudorandom number generators.

To determine the critical quantities (e.g. $\nu$ and $K_c$) with high resolution by using 
finite-size scaling analysis, it is necessary to find the peak values of
derivatives of the thermodynamic quantities and their corresponding locations with 
very high precision. As the imprecision will accumulate during calculation, 
double precision may not be enough to fulfill the task. Therefore, quadruple 
precision arithmetic has been used in the data analysis.

Additionally, we have simulated $32^3$ systems with the Wolff cluster 
flipping algorithm and the Metropolis single spin-flip algorithm. A total of $3 \times 10^{10}$ 
measurements were taken for each algorithm. The Wolff cluster simulation for $L=32$ was repeated using the MRG32K3A random number generator from Pierre L'Ecuyer, ``Combined Multiple Recursive Random Number Generators'', Operations Research, 47, 1 (1999), 159-164.  (We used the implementation by Guskova, Barash and Shchur in their rngavxlib random number library~\cite{Guskova}.)  The locations and values of the maxima in all quantities were the same, to within the error bars, as those from the Metropolis simulations and the Wolff simulations with the Mersenne Twister; and t-test comparisons yielded no p-values less than 0.2.  Hence the problems found by Ferrenberg et al.~\cite{FLWrng} using other random number generators were not noticeable here. Even though the Mersenne Twister has been tested multiple times, all computer algorithms for generating (pseudo-) random number streams will ultimately produce some small bias that will limit the accuracy of a simulation. While we have not been able to detect such effects, \textit{caveat emptor}.

\section{\label{sec:level1}results and discussion}

%%%%%%%%%%%%%%%%%%%%%%%%%%%%%%%%%%%%%%%%%%%%%%%%%%%%%%%%%%%%%%%%%%%%%%%%%%
% subsection: nu
\subsection{\label{sec:level2}Finite-size scaling analysis to determine $\nu$}
\label{fss_nu}
First, we performed an analysis with only one correction term,
\begin{equation}
X_{\max} = 
X_0 L^{1/\nu} (1 + a_1 L^{-\omega_1})
\label{fit_nu_1w}
\end{equation}
where $X$ is the quantity we have used to estimate the critical exponent 
$\nu$: the logarithmic derivatives 
${\partial \ln{\langle|m|^i\rangle}} / {\partial K}$ for $i=1,2,3,4$; 
%$\frac{\partial \ln{\langle|m|^2\rangle}}{\partial K}$, 
%$\frac{\partial \ln{\langle|m|^3\rangle}}{\partial K}$,
%$\frac{\partial \ln{\langle|m|^4\rangle}}{\partial K}$,
the magnetization cumulant derivatives 
${\partial U_{2i}} / {\partial K}$ for $i=1,2,3$.
%$\frac{\partial U_4}{\partial K}$, 
%$\frac{\partial U_6}{\partial K}$.
Least-squares fit has been performed for Eq.~(34). $\upchi^2$ per degree 
of freedom (dof) is used as the goodness of the fit, and ideally it is 
approximately 1, with values too small indicating that the error is too large and values too large indicating a poor quality of fit.  In our analysis, the $\upchi^2$ per dof is between 0.50 to 1.73 which is a reasonable range.

By calculating the covariance matrix and doing the cross-correlation analysis, 
we give estimates for $\nu$ in Table~\ref{table_nu_1w} where the minimum lattice 
size included in the analysis, $L_{\min}$, is eliminated one by one. 
\begin{table}[]
\caption {Results for the critical exponent $\nu$ when only considering one 
correction term as a function of $L_{\min}$.}
\label{table_nu_1w}
\begin{ruledtabular}
\begin{tabular}{@{\hspace{7em}}c c@{\hspace{7em}}}
$L_{\min}$ & $\nu$ \\
\colrule
16  & $0.629\,756(32)$ \\
24  & $0.629\,765(42)$ \\
32  & $0.629\,749(48)$ \\
48  & $0.630\,05(13)$ \\
64  & $0.629\,83(13)$ \\
80  & $0.629\,80(14)$ \\
96  & $0.629\,72(12)$ \\
112 & $0.629\,61(10)$ \\
128 & $0.629\,56(11)$ \\
144 & $0.629\,63(14)$ \\
160 & \hspace{0.16cm}$0.629\,554(95)$ \\
\end{tabular}
\end{ruledtabular}
\end{table}

\begin{figure} []
\centering
\includegraphics [width=0.9\hsize] {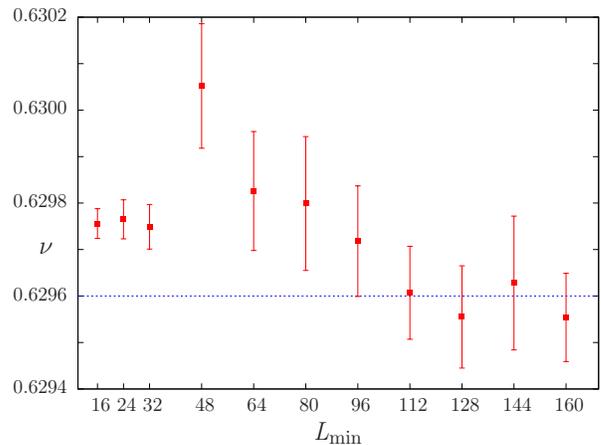}
\caption {Results for the critical exponent $\nu$ when only considering one 
correction term as a function of $L_{\min}$.}
\label{fig_nu_1w}
\end{figure}

In Fig.~\ref{fig_nu_1w}, we see that the estimated value for the 
critical exponent $\nu$ seems to be stable for small values of $L_{\min}$
($L_{\min}=16,\, 24,\, 32$). And there is a sudden jump from $L_{\min}=32$ to $L_{\min}=48$. 
Finally, $\nu$ value tends to be stable at the large lattices 
($L_{\min} \ge 112$), around $0.629\,60$.

The finite-size effect is strong when lattice sizes are small. Only 
considering one correction term is insufficient, and there is a 
systematic decrease in the value of $\nu$ as $L_{\min}$ increases. But the 
first three values for $\nu$ seem to be abnormal.  This is a consequence of a single correction term attempting to account for all finite-size 
effects with estimates for different sizes having different uncertainties.  Therefore, the value of the 
correction exponent from such fits differs from the theoretical prediction 
(0.83)~\cite{conformal_BT2014}. 
Including small lattices, the estimate for the correction exponent is larger than 
0.83 and the resulting estimate for $\nu$ is smaller than it should be. It seems to be stable at around $0.629\,75$ when $L_{\min} \le 32$. 
However, in order to minimize the least squares, all fitting 
parameters would vary altogether. As a single correction term contributes 
differently for different system sizes, it would result in inconsistent 
estimates for $\omega$ and $\nu$. In consequence, more correction terms need 
to be taken into account.

Because of the lack of a sufficient number of degrees of freedom, it is difficult to include two or
more correction terms as unknown fitting parameters. However, with the help of the 
conformal bootstrap \cite{conformal_BT2012, conformal_BT2014}, we have the 
theoretical prediction for the confluent correction exponents,
\begin{equation}
\omega_1 = 0.8303(18),\quad \omega_2 \approx 4.
\end{equation}
Additionally, we can consider the correction term corresponding to the 
non-linear scaling fields \cite{Fisher_correction},
\begin{equation}
\omega_\nu = 1 / \nu 
\end{equation}
Also, a correction term due to the rotational invariance of the lattice \cite{Massimo_wnr} may play a role,
\begin{equation}
\omega_{NR} = 2.0208(12)
\end{equation}

In our analysis we permitted any of the types of correction terms in Eq.~(\ref{dU}) and Eq.~(\ref{dlnm}) to contribute an amount that was statistically significant, but due to the finite precision of our estimates for thermodynamic quantities and the limited number of system sizes in the analysis we found that including more than three correction terms did not lead to meaningful fits. Performing least squares fits with 7 different combinations of three correction terms, yielded consistent estimates for the asymptotic values of the critical exponent $\nu$.
  
  % There may be a large number of possible correction terms, but it would introduce large statiscal errors if too many correction terms included. Although we allowed for the possibility of any of correction terms in Eq.~(\ref{dU}) and Eq.~(\ref{dlnm}) contributing an amount that was statistically significant, we found that with more than three correction terms we simply did not have a sufficient number of data points to perform meaningful fits. Performing least squares fits with 7 different combinations of three correction terms, the results for the asymptotic values of the critical exponent $\nu$ agree with each other essentially. Thus, either the higher order terms have small prefactors or we do not have sufficient resolution to differentiate their contribution from those of higher order confluent corrections.

  We have found that the best fit was obtained by using $\omega_1 = 0.83$, $\omega_2 = 4$ and $\omega_\nu = 1.6$. We will show these results in detail.

Thus, the fitting model is,
\begin{equation}
X_{\max} = 
X_0 L^{1/\nu} (1 + a_1 L^{-\omega_1} + a_2 L^{-\omega_2} + a_3 L^{-\omega_\nu})
\label{fit_nu_3w}
\end{equation}
We have considered one fixed correction exponent $\omega_1 = 0.83$, two 
fixed exponents $\omega_1 = 0.83$, $\omega_2 = 4$, and three fixed 
exponents $\omega_1 = 0.83$, $\omega_2 = 4$, $\omega_\nu = 1.6$, to the 
fitting model Eq.~(\ref{fit_nu_3w}). The results for $\nu$ are shown in 
Table~\ref{table_nu_123w_fixed}.
\begin{table}[]
\caption {Results for the critical exponent $\nu$ when considering one fixed
correction exponent $\omega_1 = 0.83$, two fixed exponents $\omega_1 = 0.83$, 
$\omega_2 = 4$, and three fixed exponents $\omega_1 = 0.83$, $\omega_2 = 4$, 
$\omega_\nu = 1.6$ as a function of $L_{\min}$.}
\label{table_nu_123w_fixed}
\begin{ruledtabular}
\begin{tabular}{c c c c}
$L_{\min}$ & $\nu $($\omega_1$ fixed) & $\nu $($\omega_{1,2}$ fixed) & 
$\nu$($\omega_{1,2,\nu}$ fixed) \\
\colrule
16  & $0.631\,814(18)$ & $0.630\,806(30)$ & $0.630\,072(45)$ \\
24  & $0.631\,046(26)$ & $0.630\,513(40)$ & $0.630\,049(57)$ \\
32  & $0.630\,722(33)$ & $0.630\,241(55)$ & $0.629\,980(77)$ \\
48  & $0.630\,350(48)$ & $0.630\,278(78)$ & $0.629\,99(11)$ \\
64  & $0.630\,319(62)$ & $0.630\,21(11)$  & $0.630\,06(15)$ \\
80  & $0.630\,285(78)$ & $0.630\,10(15)$  & $0.629\,93(21)$ \\
96  & $0.630\,25(10)$  & $0.629\,93(18)$  & $0.629\,90(29)$ \\
112 & $0.630\,14(13)$  & $0.630\,01(17)$  & $0.629\,93(18)$ \\
128 & $0.630\,04(15)$  & $0.630\,04(15)$  & $0.629\,84(22)$ \\
144 & $0.629\,85(18)$  & $0.629\,85(18)$  & $0.629\,96(26)$ \\
160 & $0.629\,95(22)$  & $0.629\,95(22)$  &  \\
\end{tabular}
\end{ruledtabular}
\end{table}

\begin{figure}[]
\centering
\includegraphics [width=0.9\hsize] {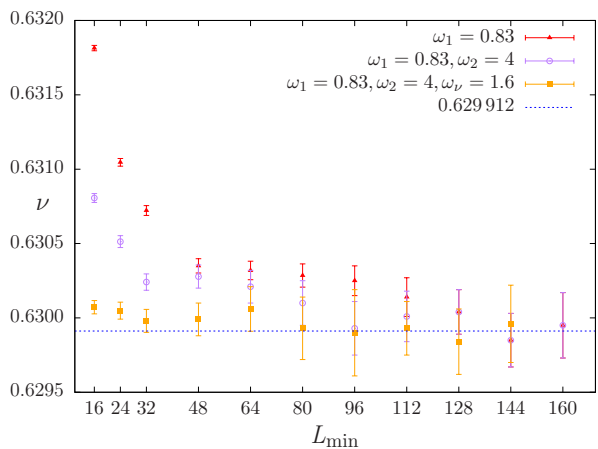}
\caption {Results for the critical exponent $\nu$ when considering one fixed
correction exponent $\omega_1 = 0.83$, two fixed exponents $\omega_1 = 0.83$, 
$\omega_2 = 4$, and three fixed exponents $\omega_1 = 0.83$, $\omega_2 = 4$, 
$\omega_\nu = 1.6$ as a function of $L_{\min}$.}
\label{fig_nu_123w_fixed}
\end{figure}
In Fig.~\ref{fig_nu_123w_fixed}, we see that, with only one 
fixed confluent correction exponent ($\omega_1 = 0.83$), the estimated value 
for the critical exponent $\nu$ decreases as $L_{\min}$ increases if 
$L_{\min} \leq 128$. The $\nu$ value seems to be stable if $L_{\min} \geq 128$. 
$\upchi^2$ per dof is very high when $L_{\min}$ is small, which indicates that 
only considering one correction term into the fit is inadequate, especially 
for the small lattice sizes ($L_{\min} = 16,\, 24,\, 32$). 
When considering two fixed confluent correction exponents 
($\omega_1 = 0.83$, $\omega_2 = 4$), $\nu$ value decreases systematically up 
to $L_{\min} = 96$. After that, the $\nu$ value appears to be statistically 
fluctuating. Still, $\upchi^2$ per dof is high when $L_{\min}$ is small, which 
means that two correction terms are not enough for small lattice sizes 
($L_{\min} = 16,\, 24$). 
Compared with the analysis with only one fixed correction 
exponent, the estimates for $\nu$ are very consistent when 
$L_{\min} \geq 128$. This is because when $L_{\min}$ becomes large enough, the 
second confluent correction term contributes little.

When considering three correction exponents, two for confluent corrections 
($\omega_1 = 0.83$, $\omega_2 = 4$) and one for the non-linear scaling fields
($\omega_\nu = 1.6$), the estimated value for the critical exponent $\nu$ 
seems to be statistically fluctuating. $\upchi^2$ per dof for each quantity 
is between 0.53 to 1.78, which is reasonable. 
But, all estimates for the critical 
exponent $\nu > 0.629\,97$ if $L_{\min} < 80$, while $\nu < 0.629\,97$ if 
$L_{\min} \geq 80$. It seems that there is still a systematic decrease 
of $\nu$ as $L_{\min}$ increases. Therefore, the value for $\nu$ is 
estimated by taking the average of $\nu$ obtained from different fits for 
different $L_{\min}$ varying from 80 to 144, 
$\nu = 0.629\,912$. To estimate the error of $\nu$, we used the
jackknife method on estimates of $\nu$ from three correction term 
analysis using different ranges of $L_{\min}$: consider estimates for $\nu$ from $L_{\min} = 96$ to 144, then do 
a jackknife analysis to estimate the value and error of $\nu$. Add one 
$\nu$ value corresponding to $L_{\min} = 80$, then do a jackknife analysis 
from $L_{\min} = 80$ to 144. Do this one by one, up to the analysis from 
$L_{\min} = 16$ to 144. Results are shown in Table~\ref{table_nu_3w_jn}.
\begin{table}
\caption {Results for the critical exponent $\nu$ from jackknife 
analysis on estimates for $\nu$ that are from the three correction terms 
analysis for values of $L_{\min}$ within the ranges shown.}
\label{table_nu_3w_jn}
\begin{ruledtabular}
\begin{tabular}{@{\hspace{7em}}c c@{\hspace{7em}}}
$L_{\min}$ & $\nu$ \\
\colrule
16-144  & $0.629\,97(21)$ \\
24-144  & $0.629\,96(19)$ \\
32-144  & $0.629\,95(16)$ \\
48-144  & $0.629\,94(16)$ \\
64-144  & $0.629\,94(15)$ \\
80-144  & \hspace{0.15cm}$0.629\,912(86)$ \\
96-144  & \hspace{0.15cm}$0.629\,908(81)$ \\
\end{tabular}
\end{ruledtabular}
\end{table}

Based on the values of $L_{\min}$ to estimate the value of $\nu$ (from 80 to 
144), we find
\begin{equation}
\nu = 0.629\,912 (86).
\label{nu_estimate}
\end{equation}

%%%%%%%%%%%%%%%%%%%%%%%%%%%%%%%%%%%%%%%%%%%%%%%%%%%%%%%%%%%%%%%%%%%%%%%%%%
% subsection: Kc
\subsection{\label{sec:level2}Finite-size scaling analysis to determine $K_c$}
\label{fss_kc}
To estimate the critical coupling $K_c$, we have considered the location of 
the peak of the logarithmic derivatives
${\partial \ln{\langle|m|^i\rangle}} / {\partial K}$ for $i=1,2,3,4$; 
the magnetization cumulant derivatives 
${\partial U_{2i}} / {\partial K}$ for $i=1,2,3$; 
the specific heat $C$; 
the derivative of the modulus of the magnetization 
${\partial \langle |m| \rangle} / {\partial K}$; 
the finite-lattice susceptibility $\chi'$; 
as well as the location of zero of the fourth-order energy cumulant $Q_4$.

First, estimate the critical coupling $K_c$ with one correction term,
\begin{equation}
K_c(L) = K_c + A_0 L^{-1/\nu} (1 + A_1 L^{-\omega_1})
\end{equation}
where the critical exponent is fixed to be $\nu = 0.629\,912$, and the 
correction exponent $\omega_1$ is unfixed. Except in the situation where 
$L_{\min} = 16$ for ${\partial \langle |m| \rangle} / {\partial K}$, the 
$\upchi^2$ per degree of freedom is high (2.76), in other cases, 
$\upchi^2$ per dof is acceptable.

By calculating the covariance matrix and doing the cross correlation analysis, 
we estimated $K_c$ as shown in Table~\ref{table_Kc_1w}. Minimum lattice 
size $L_{\min}$ that is taken into account is eliminated one by one. 
\begin{table}[]
\caption {Results for the critical coupling $K_c$ when only considering one 
correction term as a function of $L_{\min}$.}
\label{table_Kc_1w}
\begin{ruledtabular}
\begin{tabular}{@{\hspace{6em}}c c@{\hspace{6em}}}
$L_{\min}$ & $K_c$ \\
\colrule
16  & $0.221\,654\,621\,8(13)$ \\
24  & $0.221\,654\,623\,9(16)$ \\
32  & $0.221\,654\,624\,9(19)$ \\
48  & $0.221\,654\,623\,4(27)$ \\
64  & $0.221\,654\,625\,3(45)$ \\
80  & $0.221\,654\,626\,1(62)$ \\
96  & $0.221\,654\,630\,0(78)$ \\
112 & $0.221\,654\,630\,2(69)$ \\
128 & $0.221\,654\,630\,2(63)$ \\
144 & \hspace{-0.2cm}$0.221\,654\,628(13)$ \\
160 & \hspace{0.02cm}$0.221\,654\,630\,3(85)$ \\
\end{tabular}
\end{ruledtabular}
\end{table}

\begin{table*} 
\caption {Results for the critical coupling $K_c$ from fits with: (left column) a single correction term (fixed correction exponent $\omega_1 = 0.83$); (center column) two correction terms (fixed exponents $\omega_1 = 0.83$, 
$\omega_2 = 4$; and (right column) three correction terms (fixed exponents $\omega_1 = 0.83$, $\omega_2 = 4$, 
$\omega_\nu = 1.6$) as a function of $L_{\min}$.}
\label{table_Kc_123w_fixed}
\begin{ruledtabular}
\begin{tabular}{@{\hspace{8em}}c c c c@{\hspace{8em}}}
$L_{\min}$ & $K_c$(1 fixed $\omega$) & $K_c$(2 fixed $\omega$) & 
$K_c$(3 fixed $\omega$) \\
\colrule
16  & $0.221\,654\,656\,2(10)$ & $0.221\,654\,639\,3(11)$ & $0.221\,654\,625\,7(21)$ \\
24  & $0.221\,654\,638\,8(11)$ & $0.221\,654\,630\,8(12)$ & $0.221\,654\,625\,7(24)$ \\
32  & $0.221\,654\,634\,3(11)$ & $0.221\,654\,630\,7(12)$ & $0.221\,654\,625\,3(32)$ \\
48  & $0.221\,654\,630\,7(12)$ & $0.221\,654\,630\,5(12)$ & $0.221\,654\,623\,2(30)$ \\
64  & $0.221\,654\,628\,4(13)$ & $0.221\,654\,628\,4(13)$ & $0.221\,654\,623\,4(60)$ \\
80  & $0.221\,654\,627\,5(14)$ & $0.221\,654\,627\,5(15)$ & $0.221\,654\,625\,0(75)$ \\
96  & $0.221\,654\,626\,0(17)$ & $0.221\,654\,626\,0(16)$ & $0.221\,654\,627\,9(97)$ \\
112 & $0.221\,654\,625\,9(18)$ & $0.221\,654\,626\,0(18)$ & $0.221\,654\,625\,0(49)$ \\
128 & $0.221\,654\,625\,8(21)$ & $0.221\,654\,625\,8(21)$ & $0.221\,654\,626\,3(48)$ \\
144 & $0.221\,654\,627\,0(25)$ & $0.221\,654\,627\,0(25)$ & $0.221\,654\,627\,1(34)$ \\
160 & $0.221\,654\,626\,3(23)$ & $0.221\,654\,626\,4(23)$ &  \\
\end{tabular}
\end{ruledtabular}
\end{table*}

\begin{figure} []
\centering
\includegraphics [width=0.95\hsize] {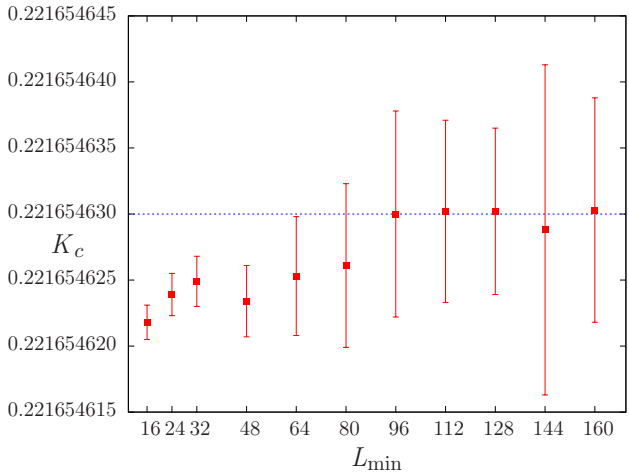}
\caption {Results for the critical coupling $K_c$ with only one 
correction term included in the fitting as a function of $L_{\min}$.}
\label{fig_Kc_1w}
\end{figure}

In Fig.~\ref{fig_Kc_1w}, we can see that, the estimated value for the 
critical coupling $K_c$ appears to be stable if $L_{\min} \geq 96$, around 
$0.221\,654\,630$. 

Similar to the analysis to determine $\nu$ we used seven different combinations of the three correction terms and found that the choice had negligible impact on the estimate for $K_c$. The best fit was obtained by using $\omega_1 = 0.83$, $\omega_2 = 4$ and $\omega_\nu = 1.6$. We will show these results in detail.
  %Likewise, 7 different models (correction combinations) were used but the results were rather insensitive to the choice of the model.

With the help of the theoretical prediction, we have considered one 
fixed correction exponent $\omega_1 = 0.83$, two fixed exponents 
$\omega_1 = 0.83$, $\omega_2 = 4$, and three fixed exponents 
$\omega_1 = 0.83$, $\omega_2 = 4$, $\omega_\nu = 1.6$, to the fitting model 
Eq.~(\ref{Kc_3w}). 
\begin{equation}
K_c(L) = K_c + A_0 L^{-1/\nu} (1 + A_1 L^{-\omega_1} + A_2 L^{-\omega_2} 
+ A_3 L^{-\omega_\nu})
\label{Kc_3w}
\end{equation}
The results for $K_c$ are shown in Table~\ref{table_Kc_123w_fixed}.

\begin{figure} []
\centering
\includegraphics [width=0.95\hsize] {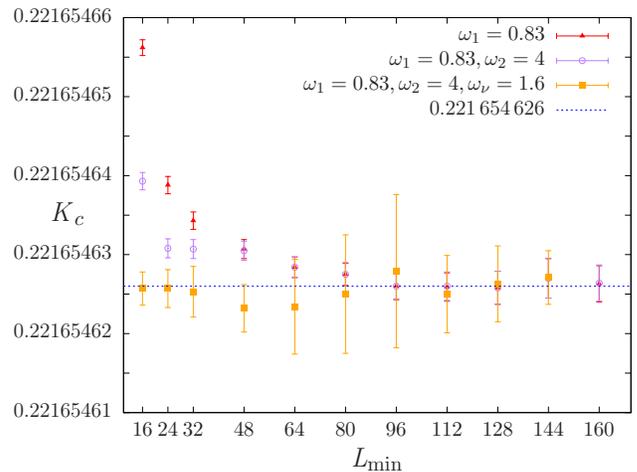}
\caption {Results for the critical coupling $K_c$ when considering one fixed
correction exponent $\omega_1 = 0.83$, two fixed exponents $\omega_1 = 0.83$, 
$\omega_2 = 4$, and three fixed exponents $\omega_1 = 0.83$, $\omega_2 = 4$, 
$\omega_\nu = 1.6$ as a function of $L_{\min}$.}
\label{fig_Kc_123w_fixed}
\end{figure}
In Fig.~\ref{fig_Kc_123w_fixed}, we can see that, when considering only one 
fixed confluent correction exponent ($\omega_1 = 0.83$), the estimated value 
for the critical coupling $K_c$ decreases as $L_{\min}$ increases if 
$L_{\min} \leq 80$. The $K_c$ value appears to be stable if $L_{\min} \geq 80$. 
$\upchi^2$ per dof is very high when $L_{\min}$ is small, which means that the 
quality of the fit is not good with one correction term when the lattice 
size is small ($L_{\min} = 16,\,24,\,32$).
When considering two fixed confluent correction exponents 
($\omega_1 = 0.83$, $\omega_2 = 4$), the $K_c$ value decreases systematically up 
to $L_{\min} = 80$ as well. After that, the $K_c$ value appears to be statistically 
fluctuating. Still, $\upchi^2$ per dof is high when $L_{\min}$ is small, which 
indicates that two correction terms are not enough for small lattice sizes 
($L_{\min} = 16,\, 24$). Compared with the analysis with only one fixed 
correction exponent, the estimates for $K_c$ are highly consistent when 
$L_{\min} \geq 64$. This is because when $L_{\min}$ becomes large enough, the 
second confluent correction term contributes little, and these two analyses 
tend to generate similar results. 

When considering three correction exponents, two for confluent corrections 
($\omega_1 = 0.83$, $\omega_2 = 4$) and one for non-linear scaling fields 
($\omega_\nu = 1.6$), $\upchi^2$ per dof for each quantity is decent except 
the following cases: 

$\upchi^2$ per dof = 2.52, if $L_{\min} = 144$ for 
${\partial \ln{\langle|m|\rangle}} / {\partial K}$, 

$\upchi^2$ per dof = 2.41, if $L_{\min} = 144$ for 
${\partial \ln{\langle|m|^2\rangle}} / {\partial K}$,
 
$\upchi^2$ per dof = 2.34, if $L_{\min} = 144$ for $\chi'$, 

$\upchi^2$ per dof = 2.89, if $L_{\min} = 144$ for 
${\partial U_{4}} / {\partial K}$,  

$\upchi^2$ per dof = 2.84, if $L_{\min} = 144$ for 
${\partial U_{6}} / {\partial K}$. 

\noindent This is because of the lack of degrees of freedom when $L_{\min}$ is large. 

The estimated value for the critical coupling $K_c$ 
appears to be statistically fluctuating. The fluctuation of $K_c$ when 
$L_{\min} \leq 80$ is larger than the one when $L_{\min} \geq 80$. Additionally, 
finite-size effect reduces as larger lattice sizes are considered. Thus, 
the value of $K_c$ is estimated through the average of $K_c$ for 
$L_{\min} =$ 80 to 144, which is $0.221\,654\,626\,2$. Likewise, a jackknife 
analysis has been done on the estimates for $K_c$ which are obtained from the three 
correction terms analysis. Results are shown in Table~\ref{table_Kc_3w_jn}.
\begin{table} []
\caption {Results for the critical coupling $K_c$ from jackknife 
analysis on estimates for $K_c$ that are from the three correction terms 
analysis.}
\label{table_Kc_3w_jn}
\begin{ruledtabular}
\begin{tabular}{@{\hspace{6em}}c c@{\hspace{6em}}}
$L_{\min}$ & $K_c$ \\
\colrule
16-144  & $0.221\,654\,625\,5(42)$ \\
24-144  & $0.221\,654\,625\,4(41)$ \\
32-144  & $0.221\,654\,625\,4(41)$ \\
48-144  & $0.221\,654\,625\,4(40)$ \\
64-144  & $0.221\,654\,625\,8(33)$ \\
80-144  & $0.221\,654\,626\,2(23)$ \\
96-144  & $0.221\,654\,626\,6(18)$ \\
\end{tabular}
\end{ruledtabular}
\end{table}

Based on the values of $L_{\min}$ from 80 to 
144 we estimate
$K_c = 0.221\,654\,626\,2 (23)$.whereas using
 $L_{\min} =$ 16 to 144, the estimate 
for the critical coupling would be,
$K_c = 0.221\,654\,625\,5 (42)$.
Therefore, our final estimate from the finite size scaling analysis, with conservative error bars, is
\begin{equation}
K_c = 0.221\,654\,626 (5).
\end{equation} 

%%%%%%%%%%%%%%%%%%%%%%%%%%%%%%%%%%%%%%%%%%%%%%%%%%%%%%%%%%%%%%%%%%%%%%%%%%
% subsection: crossing technique
\subsection{\label{sec:level2}Crossing technique of the 4th order magnetization cumulant}
\label{crossing_kc}
\begin{figure} []
\centering
\includegraphics [width=0.9\hsize] {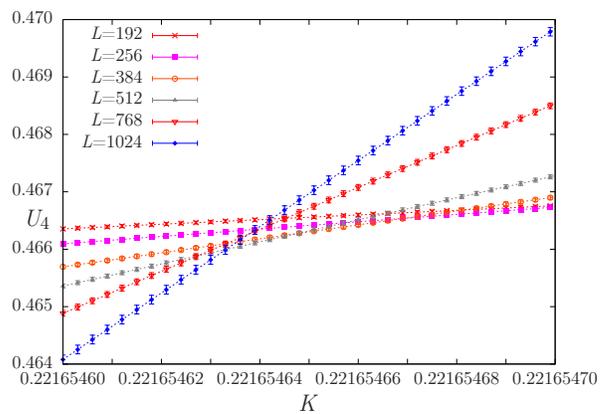}
\caption {Inverse temperature $K$ dependence of the fourth order 
magnetization cumulant $U_4$ for $L \times L \times L$ Ising lattices.}
\label{fig_U4-K}
\end{figure}

As the lattice size $L \rightarrow \infty$, the fourth-order magnetization 
cumulant $U_4 \rightarrow 0$ for $K < K_c$ and $U_4 \rightarrow 2/3$ for 
$K > K_c$. For large enough lattice sizes, curves for $U_4$ cross as a 
function of inverse temperature at a ``fixed point'' $U^*$, and the location 
of the crossing ``fixed point'' is $K_c$. 
Because the lattices are not infinitely large, finite-size 
 correction terms will prevent all curves from crossing at a common 
intersection (as in Fig.~\ref{fig_U4-K}). However, Fig.~\ref{fig_U4-K} gives 
us a preliminary estimate for $K_c$.

\begin{table} []
\caption {Results for the critical coupling $K_c$ obtained using the cumulant crossing 
technique with one correction term.}
\label{table_Kcross_1w}
\begin{ruledtabular}
\begin{tabular}{c c c c}
$L_{\min}$ & $K_c$ & dof & $\upchi^2$ per dof\\
\colrule
16   & $0.221\,654\,628\,72(41)$ & 131 & 1.64 \\
24   & $0.221\,654\,626\,85(50)$ & 115 & 1.10 \\
32   & $0.221\,654\,626\,17(58)$ & 100 & 1.06 \\
48   & $0.221\,654\,624\,63(75)$ & 86  & 0.88 \\
64   & $0.221\,654\,625\,44(91)$ & 73  & 0.89 \\
80   & $0.221\,654\,626\,3(11)$  & 61  & 0.90 \\
96   & $0.221\,654\,627\,7(12)$  & 50  & 0.84 \\
112  & $0.221\,654\,628\,0(15)$  & 40  & 0.93 \\
128  & $0.221\,654\,628\,4(17)$  & 31  & 1.04 \\
144  & $0.221\,654\,627\,8(22)$  & 23  & 1.18 \\
160  & $0.221\,654\,629\,3(24)$  & 16  & 1.29 \\
192  & $0.221\,654\,629\,5(33)$  & 10  & 1.70 \\
\end{tabular}
\end{ruledtabular}
\end{table}
The locations of the cumulant crossings have been fitted to 
Eq.~(\ref{eq_Kcross}) with one correction term. All of the parameters are 
allowed to vary independently, i.e., no fixed values for $\nu$ and $\omega$. 
Results are shown in Table~\ref{table_Kcross_1w}, where $L_{\min}$ is the 
minimum lattice size taken into account.

Additionally, the locations of the cumulant crossings have been fitted to 
Eq.~(\ref{eq_Kcross}) with two correction terms. Results are shown in 
Table~\ref{table_Kcross_2w}.
\begin{table}[]
\caption {Results for the critical coupling $K_c$ by using cumulant crossing 
technique with two correction terms.}
\label{table_Kcross_2w}
\begin{ruledtabular}
\begin{tabular}{c c c c}
$L_{\min}$ & $K_c$ & dof & $\upchi^2$ per dof\\
\colrule
16   & $0.221\,654\,624\,83(95)$ & 129 & 0.94 \\
24   & \hspace{-0.24cm} $0.221\,654\,624\,9(10)$  & 113 & 0.96 \\
32   & $0.221\,654\,624\,50(80)$ & 98  & 0.85 \\
48   & $0.221\,654\,624\,63(85)$ & 84  & 0.90 \\
64   & \hspace{-0.24cm} $0.221\,654\,625\,4(10)$  & 71  & 0.91 \\
\end{tabular}
\end{ruledtabular}
\end{table}
For $L_{\min} > 24$, the second correction term is ill-defined, and by 
$L_{\min} = 80$, the calculation gives identical values for the two 
correction exponents. This is because we lack precision to include 
two correction terms for the crossing technique.

\begin{figure} []
\centering
\includegraphics [width=0.95\hsize] {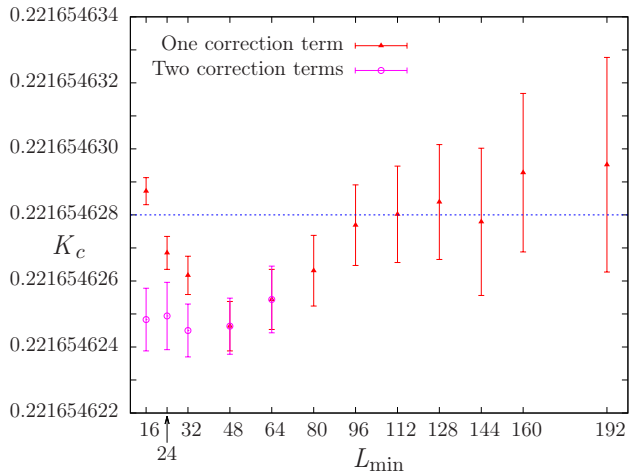}
\caption {Results for the critical coupling $K_c$ using cumulant crossings 
with one correction term and two correction terms.}
\label{fig_Kcross}
\end{figure}

In Fig.~\ref{fig_Kcross}, the critical coupling appears to be stable if 
$L_{\min} \geq 96$. 
The value of $K_c$ can be estimated by taking the average of $K_c$ values 
for $L_{\min} \geq 96$, which is $0.221\,654\,628\,4$. 
A jackknife analysis has been done on the estimates for $K_c$ that are from 
the one correction term analysis. Results are shown in 
Table~\ref{table_Kcross_jn}. 
\begin{table}[]
\caption {Results for the critical coupling $K_c$ by using jackknife 
analysis on estimates for $K_c$ that are from the cumulant crossing 
technique with one correction term analysis.}
\label{table_Kcross_jn}
\begin{ruledtabular}
\begin{tabular}{@{\hspace{6em}}c c@{\hspace{6em}}}
$L_{\min}$ & $K_c$ \\
\colrule
16-144  & $0.221\,654\,627\,4(49)$ \\
24-144  & $0.221\,654\,627\,3(47)$ \\
32-144  & $0.221\,654\,627\,3(46)$ \\
48-144  & $0.221\,654\,627\,5(45)$ \\
64-144  & $0.221\,654\,627\,8(34)$ \\
80-144  & $0.221\,654\,628\,1(24)$ \\
96-144  & $0.221\,654\,628\,4(16)$ \\
112-144 & $0.221\,654\,628\,6(14)$ \\
128-144 & $0.221\,654\,628\,7(12)$ \\
\end{tabular}
\end{ruledtabular}
\end{table}

Using results for $L_{\min}$ (96 to 192) we estimate 
\begin{equation}
K_c = 0.221\,654\,628(2)
\end{equation}

%%%%%%%%%%%%%%%%%%%%%%%%%%%%%%%%%%%%%%%%%%%%%%%%%%%%%%%%%%%%%%%%%%%%%%%%%%
% subsection: Alternative finite-size scaling analysis
\subsection{\label{sec:level2}Alternative finite-size scaling analysis}
\label{alter_fss}
In Sec.~\ref{fss_nu}, a finite-size scaling analysis was performed by looking at the magnitude of quantities at the peak locations. Alternatively,
critical exponents can be estimated by looking at quantities at our estimate for $K_c$ 
(denoted $K_{c}^{est} = 0.221\,654\,626$, i.e. the 
estimated value for $K_c$ for an infinite lattice). 
\begin{equation}
  X(K = K_{c}^{est}) = X_{0} L^{\lambda} (1 + a_1 L^{-\omega_1} + \cdots),
  \label{alternative_FSS}
\end{equation}
where $X$ is the quantity being used to determine the critical exponent $\lambda$.  For the susceptibility and the specific heat Eq.~(\ref{alternative_FSS}) includes an analytic background term.
  %There is an analytic background term in Eq.~(\ref{alternative_FSS}) for the susceptibility and the specifit heat.}

$\nu$ can be estimated from derivatives of magnetization cumulants and 
logarithmic derivatives of the magnetization at $K_{c}^{est}$. By doing the fit 
with three fixed correction exponents, and by calculating the jackknife 
covariance matrix and doing the cross correlation analysis, we find $\nu$ to be 
\begin{equation}
\nu = 0.629\,93(10).
\end{equation}
This result agrees with the value of $\nu$ estimated from Eq.~(\ref{nu_estimate}).

By examining the scaling behavior of 
the susceptibility at $K_{c}^{est}$, we have found 
that $\gamma / \nu = 1.963\,90(45)$ . 
Combining this value with our estimate for $\nu$ at Eq.~(\ref{nu_estimate}), and
assuming that exponent estimates for $\gamma$ and $\nu$ are independent, we have determined the critical exponent $\gamma$ of the magnetic susceptibility to be
\begin{equation}
\gamma = 1.237\,08(33).
\end{equation}

We also performed an analysis of the susceptibility at constant $U_4$ as suggested by Hasenbusch~\cite{Hasenbusch2010}. Fixing $U_4 = 0.4655$ and including the higher order confluent corrections to scaling we found that $\gamma = 1.237\,01(28)$, a value that is almost identical to, and with only a slightly smaller error bar than, the value obtained from finite size scaling of the susceptibility.

Because of the large analytic background in the specific heat (see Eq.~(\ref{specific_heat_scaling})), it was not possible to extract estimates of the exponent $\alpha$ with comparable precision to the other exponents evaluated here.  For this reason, we have not quoted an estimated value.

Similarly, by considering the critical behavior of $|m|$ at $K_{c}^{est}$, 
we obtained $\beta / \nu = 0.518\,01(35)$, or 
\begin{equation}
\beta = 0.326\,30(22). 
\end{equation}

%%%%%%%%%%%%%%%%%%%%%%%%%%%%%%%%%%%%%%%%%%%%%%%%%%%%%%%%%%%%%%%%%%%%%%%%%%
% subsection: Self-consistency check
\subsection{\label{sec:level2}Self-consistency check}
Inspired by a recent 3d bond and site percolation study \cite{deng_percolation}, a noticeable off-critical behavior would be observed when Monte Carlo data are 3 error bars away from the critical point. 

Following is the cumulant's ansatz \cite{cumulant_Binder1981},
\begin{equation}
	U_4(L) = U^* (1 + c L^{-\omega_1})
	\label{U4_ansatz}
\end{equation}
where $U_4$ is the 4th order cumulant and $U^*$ is a "fixed point". 

\begin{figure} []
\centering
\includegraphics [width=0.95\hsize] {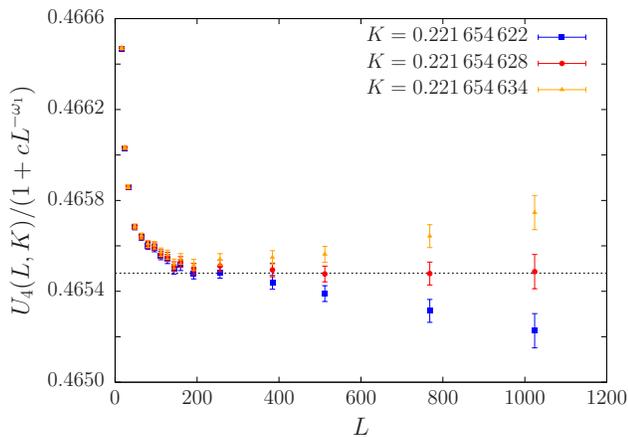}
\caption {Plot of the 4th order magnetization cumulant as a function of $L$ for fixed $K$ values. The value of $c$ was estimated by doing a fit for $U_4$ by Eq.~(\ref{U4_ansatz}). The dashed line indicates our asymptotic value for $U^*$.}
\label{fig_U4_1w}
\end{figure}

To justify our quoted error bars for the crossing technique, $K_c = 0.221\,654\,628(2)$, we performed a plot of 4th order magnetization cumulant at $K = 0.221\,654\,622$, $0.221\,654\,628$ and $0.221\,654\,634$ in Fig.~\ref{fig_U4_1w}. The value of $c$ was estimated by doing a fit for the cumulant by Eq.~(\ref{U4_ansatz}). It was generated at the estimated critical inverse temperature, with a fixed correction exponent $\omega_1 = 0.83$, over the range of $L = 144$ to 1024. It can be seen that, the data at $K = 0.221\,654\,622$ and $K = 0.221\,654\,634$ begin to diverge as $L$ increases, while the data at $K = 0.221\,654\,628$ converge to $U^* = 0.465\,48(5)$. Our estimate is consistent with $0.465\,45(13)$ from Bl\"ote et al~\cite{cluster_processor}, but higher than $0.465\,306(34)$ from Deng and Bl\"ote~\cite{Blote_Deng_2003}.

\begin{figure} []
\centering
\includegraphics [width=0.95\hsize] {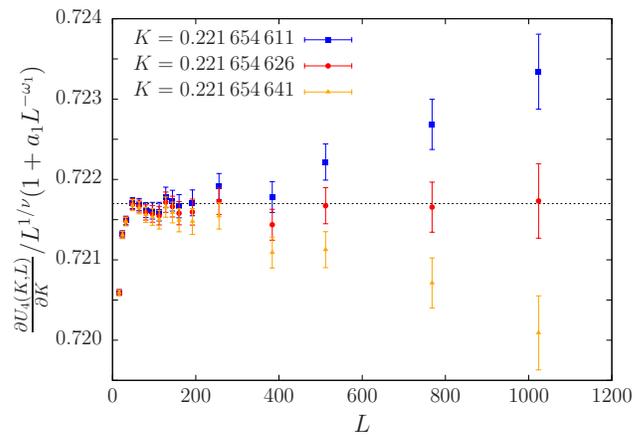}
\caption {Plot of the derivative the 4th order magnetization cumulant as a function of $L$ for fixed $K$ values. The value of $a_1$ was estimated by doing a fit for ${\partial U_{4}} / {\partial K}$ by Eq.~(\ref{alternative_FSS}).}
\label{fig_dU4_1w}
\end{figure}

Similarly, a plot of the derivative of the 4th order magnetization cumulant is shown in Fig.~\ref{fig_dU4_1w}. Based on the FSS estimate $K_c = 0.221\,654\,626 (5)$ in Sec.~\ref{fss_kc}, the data away from the estimated critical point by 3 error bars have a noticeable divergence. 

All in all, Fig.~\ref{fig_U4_1w} and Fig.~\ref{fig_dU4_1w} indicate that our quoted error bars for $K_c$ from the crossing technique and the FSS are reliable.

%%%%%%%%%%%%%%%%%%%%%%%%%%%%%%%%%%%%%%%%%%%%%%%%%%%%%%%%%%%%%%%%%%%%%%%%%%
% subsection: Discussion

\begin{table*} []
\caption {Comparison of our results for the critical coupling $K_c$ and the critical exponents $\nu$, $\gamma$ with other recently obtained values. The number marked with * is not given by the reference directly, but is calculated by Fisher's scaling law $\gamma = \nu (2-\eta)$. The error is calculated using simple error propagation, which assumes that $\nu$ and $\eta$ are independent and uncorrelated. \\ 
$\rm{^a}$ Special purpose computer. \\
$\rm{^b}$ Monte Carlo study of the non-linear relaxation function. }
\label{table_comparison}
\begin{ruledtabular}
\begin{tabular}{c c l l l}
Reference & Method & \multicolumn{1}{c}{$K_c$} 
& \multicolumn{1}{c}{$\nu$} & \multicolumn{1}{c}{$\gamma$} \\
\colrule
%El-Showk et al.(2014)~\cite{conformal_BT2014} & conformal bootstrap
%& & 0.629\,99(5) & 1.237\,1(1)* \\
Butera and Comi(2002)~\cite{HT_expan2002} & HT series
& $0.221\,655(2)$ & 0.629\,9(2) & 1.237\,1(1) \\

Bl\"ote et al.(1999)~\cite{cluster_processor}$\rm{^a}$ & MC
& $0.221\,654\,59(10)$ & 0.630\,32(56) & 1.237\,2(13)* \\

Deng and Bl\"ote(2003)~\cite{Blote_Deng_2003} & MC
& $0.221\,654\,55(3)$ & 0.630\,20(12) & 1.237\,2(4)* \\

Ozeki and Ito(2007)~\cite{Ito}$\rm{^b}$ & MC NL relax
& $0.221\,654\,7(5)$ & 0.635(5) & 1.255(18)* \\

Weigel and Janke(2010)~\cite{cross_correlation2010} & MC
& $0.221\,657\,03(85)$ & 0.630\,0(17) & 1.240\,9(62)* \\
Hasenbusch(2010)~\cite{Hasenbusch2010} & MC
& $0.221\,654\,63(8)$ & 0.630\,02(10) & 1.237\,19(21)* \\
Kaupu\~zs(2011)~\cite{MC_non_per1} & MC
& $0.221\,654\,604(18)$ &  & \\

Kos et al.(2016)~\cite{conformal_BT2016} & conformal bootstrap
& & 0.629\,971(4) & 1.237\,075(8)* \\

Wang et al.(2014)~\cite{tensor_RG_HO} & tensor RG 
& $ 0.221\,654\,555\,5(5)$  &  &  \\
Rosengren(1986)~\cite{Rosengren_exact} & conjecture 
& $0.221\,658\,63\cdots$ & & \\
%Zhang(2007)~\cite{Zhang_exact} & conjecture 
%& $0.240\,605\,91\cdots$ & $0.666\,666\cdots$ & $1.25$ \\
\hspace{0.7cm}Our results (no fit assumptions) & MC & 0.221\,654\,630(7)  & 0.629\,60(15) & 1.236\,41(45) \\
\hspace{0.26cm}Our results (constrained fits) & MC & $0.221\,654\,626(5)$ & 0.629\,912(86) & 1.237\,08(33) \\
\hspace{0.76cm}Our results (cumulant crossings)& MC & $0.221\,654\,628(2)$ & & \\
\hspace{-0.23cm}Our results (constant $U_4$) & MC & & & 1.237\,01(28) \\
\end{tabular}
\end{ruledtabular}
\end{table*}

\subsection{\label{sec:level2}Discussion}
It is only because of the combination of an efficient, cluster-flipping Monte Carlo algorithm, high statistics simulations, histogram reweighting, and a cross-correlation jackknife analysis that we were able to achieve the high resolution results presented earlier in this Section.  Now, we can compare our estimates for $K_c$ and $\nu$ with other high-resolution result from simulation and theory. 
Table~\ref{table_comparison} shows the comparison.

In Sec.~\ref{fss_nu}, we determined the critical exponent of the 
correlation length $\nu = 0.629\,912(86)$. Our value is perfectly consistent (i.e. within the error bars) with the recent conformal bootstrap result of 
Kos et al.~\cite{conformal_BT2016}, as well as that from an older work by 
El-Showk et al.~\cite{conformal_BT2014}. In addition, our result agrees with the high-temperature result of Butera and Comi~\cite{HT_expan2002}, Monte Carlo result of Deng and Bl\"ote~\cite{Blote_Deng_2003}, and nonequilibrium relaxation Monte Carlo result of Ozeki and Ito~\cite{Ito}. Also, our result agrees well with 
the Monte Carlo result of Hasenbusch~\cite{Hasenbusch2010} but is lower than that of
Weigel and Janke~\cite{cross_correlation2010}; however, within the respective error bars there is agreement although we have substantially higher precision than either of these previous studies.  Our system sizes and statistics are substantially greater than those used by Weigel and Janke, and Hasenbusch examined the behavior of the ratio of partition functions $Z_a /Z_p$, 
and the second moment correlation length over the linear
lattice size $\xi_2/ L$ so the methodologies are not identical.  Our estimate for $K_c$ differs from that obtained by Kaupu\~zs et al~\cite{MC_non_per1} using a parallel Wolff algorithm by an amount that barely agrees to within the error bars.  Somewhat perplexingly, they were able to fit their data to two rather different values of $\nu$, so no comparison of critical exponents is possible.

The recent tensor 
renormalization group result for $K_c$~\cite{tensor_RG_HO} does not agree 
with our result; in fact the difference is many times the respective error bars. 

To place these results in perspective, it is interesting to note that as far back as 1982 Gaunt's high temperature series expansions~\cite{Gaunt_1982} gave the estimate $K_c = 0.221\,66(1)$ and in 1983 Adler~\cite{Adler_1983} estimated $0.221\,655 < K_c < 0.221\,656$ with confluent corrections included in the analysis.

Neither the Rosengren's ``exact conjecture'' nor Zhang's 
so-called ``exact'' solution agree with our numerical values, thus adding further evidence to the already strong arguments that neither are, in fact, exact.

In Sec.~\ref{alter_fss}, we have estimated the critical exponents by using an 
alternative finite-size scaling analysis. The critical exponent of the 
correlation length is estimated to be $\nu = 0.629\,93(10)$, which is 
consistent with our estimate in Sec.~\ref{fss_nu}.  While our final estimate is slightly lower than the best alternative values, there is agreement to within the error bars. Also, our estimate $\gamma = 1.237\,08(33)$ is consistent with the conformal bootstrap estimates given by Kos et al.~\cite{conformal_BT2016}, El-Showk et al.~\cite{conformal_BT2014}, and slightly smaller than the Monte Carlo estimates by Deng and Bl\"ote~\cite{Blote_Deng_2003}, Hasenbusch~\cite{Hasenbusch2010}, 
and Weigel and Janke~\cite{cross_correlation2010}; but, once again, there is overlap within the respective error bars.

%%%%%%%%%%%%%%%%%%%%%%%%%%%%%%%%%%%%%%%%%%%%%%%%%%%%%%%%%%%%%%%%%%%%%%%%%%
\section{\label{sec:level1}conclusion}
We have studied a 3d Ising model with the Wolff cluster flipping algorithm, histogram reweighting, and finite size scaling including cross-correlations using quadruple precision arithmetic for the analysis.  Using a wide range of system sizes, with the largest containing more than $10^9$ spins, and including corrections to scaling, we have obtained results for $K_c$, $\nu$, and $\gamma$ that are comparable in precision to those from the latest theoretical predictions and can provide independent verification of the predictions from those methods.  Our values provide further numerical evidence that none of the purported ``exact'' values are correct.  To within error bars we obtain the same value for the critical exponent $\nu$ as that predicted by the conformal bootstrap; however, our estimate for the critical temperature $K_c$ does not agree with the result from the tensor renormalization group to within the respective error bars.

As efforts to increase Monte Carlo precision continue, new sources of error must be taken into account. Future attempts to substantially improve precision will need to carry out more stringent tests of the random number generator and acquire much greater statistics for much larger lattice sizes.  Such simulations and subsequent analysis would require orders of magnitude greater computer resources and would thus be non-trivial.

%%%%%%%%%%%%%%%%%%%%%%%%%%%%%%%%%%%%%%%%%%%%%%%%%%%%%%%%%%%%%%%%%%%%%%%%%%
\vspace{1.0cm}
\begin{acknowledgments}
We thank Dr. M. Weigel and Dr. S.-H. Tsai for valuable discussions. Computing resources were provided by the Georgia Advanced Computing Resource Center, the Ohio Supercomputing Center, and the Miami University Computer Center.
\end{acknowledgments}

%\nocite{*}

\bibliography{3dising}% Produces the bibliography via BibTeX.

\begin{thebibliography}{45}%
\makeatletter
\providecommand \@ifxundefined [1]{%
 \@ifx{#1\undefined}
}%
\providecommand \@ifnum [1]{%
 \ifnum #1\expandafter \@firstoftwo
 \else \expandafter \@secondoftwo
 \fi
}%
\providecommand \@ifx [1]{%
 \ifx #1\expandafter \@firstoftwo
 \else \expandafter \@secondoftwo
 \fi
}%
\providecommand \natexlab [1]{#1}%
\providecommand \enquote  [1]{``#1''}%
\providecommand \bibnamefont  [1]{#1}%
\providecommand \bibfnamefont [1]{#1}%
\providecommand \citenamefont [1]{#1}%
\providecommand \href@noop [0]{\@secondoftwo}%
\providecommand \href [0]{\begingroup \@sanitize@url \@href}%
\providecommand \@href[1]{\@@startlink{#1}\@@href}%
\providecommand \@@href[1]{\endgroup#1\@@endlink}%
\providecommand \@sanitize@url [0]{\catcode `\\12\catcode `\$12\catcode
  `\&12\catcode `\#12\catcode `\^12\catcode `\_12\catcode `\%12\relax}%
\providecommand \@@startlink[1]{}%
\providecommand \@@endlink[0]{}%
\providecommand \url  [0]{\begingroup\@sanitize@url \@url }%
\providecommand \@url [1]{\endgroup\@href {#1}{\urlprefix }}%
\providecommand \urlprefix  [0]{URL }%
\providecommand \Eprint [0]{\href }%
\providecommand \doibase [0]{http://dx.doi.org/}%
\providecommand \selectlanguage [0]{\@gobble}%
\providecommand \bibinfo  [0]{\@secondoftwo}%
\providecommand \bibfield  [0]{\@secondoftwo}%
\providecommand \translation [1]{[#1]}%
\providecommand \BibitemOpen [0]{}%
\providecommand \bibitemStop [0]{}%
\providecommand \bibitemNoStop [0]{.\EOS\space}%
\providecommand \EOS [0]{\spacefactor3000\relax}%
\providecommand \BibitemShut  [1]{\csname bibitem#1\endcsname}%
\let\auto@bib@innerbib\@empty
%</preamble>
\bibitem [{\citenamefont {Ising}(1925)}]{Ising}%
  \BibitemOpen
  \bibfield  {author} {\bibinfo {author} {\bibfnamefont {E.}~\bibnamefont
  {Ising}},\ }\href@noop {} {\bibfield  {journal} {\bibinfo  {journal} {Z.
  Phys.}\ }\textbf {\bibinfo {volume} {31}},\ \bibinfo {pages} {253} (\bibinfo
  {year} {1925})}\BibitemShut {NoStop}%
\bibitem [{\citenamefont {Onsager}(1944)}]{Onsager}%
  \BibitemOpen
  \bibfield  {author} {\bibinfo {author} {\bibfnamefont {L.}~\bibnamefont
  {Onsager}},\ }\href@noop {} {\bibfield  {journal} {\bibinfo  {journal} {Phys.
  Rev.}\ }\textbf {\bibinfo {volume} {65}},\ \bibinfo {pages} {117} (\bibinfo
  {year} {1944})}\BibitemShut {NoStop}%
\bibitem [{\citenamefont {Ferrenberg}\ and\ \citenamefont
  {Landau}(1991)}]{Ferrenberg1991}%
  \BibitemOpen
  \bibfield  {author} {\bibinfo {author} {\bibfnamefont {A.~M.}\ \bibnamefont
  {Ferrenberg}}\ and\ \bibinfo {author} {\bibfnamefont {D.~P.}\ \bibnamefont
  {Landau}},\ }\href@noop {} {\bibfield  {journal} {\bibinfo  {journal} {Phys.
  Rev. B}\ }\textbf {\bibinfo {volume} {44}},\ \bibinfo {pages} {5081}
  (\bibinfo {year} {1991})}\BibitemShut {NoStop}%
\bibitem [{\citenamefont {Ozeki}\ and\ \citenamefont {Ito}(2007)}]{Ito}%
  \BibitemOpen
  \bibfield  {author} {\bibinfo {author} {\bibfnamefont {Y.}~\bibnamefont
  {Ozeki}}\ and\ \bibinfo {author} {\bibfnamefont {N.}~\bibnamefont {Ito}},\
  }\href@noop {} {\bibfield  {journal} {\bibinfo  {journal} {J. Phys. A: Math.
  Theor.}\ }\textbf {\bibinfo {volume} {40}},\ \bibinfo {pages} {R149}
  (\bibinfo {year} {2007})}\BibitemShut {NoStop}%
\bibitem [{\citenamefont {Bl\"ote}\ \emph {et~al.}(1996)\citenamefont
  {Bl\"ote}, \citenamefont {Heringa}, \citenamefont {Hoogland}, \citenamefont
  {Meyer},\ and\ \citenamefont {Smit}}]{blote1996}%
  \BibitemOpen
  \bibfield  {author} {\bibinfo {author} {\bibfnamefont {H.~W.~J.}\
  \bibnamefont {Bl\"ote}}, \bibinfo {author} {\bibfnamefont {J.~R.}\
  \bibnamefont {Heringa}}, \bibinfo {author} {\bibfnamefont {A.}~\bibnamefont
  {Hoogland}}, \bibinfo {author} {\bibfnamefont {E.~W.}\ \bibnamefont {Meyer}},
  \ and\ \bibinfo {author} {\bibfnamefont {T.~S.}\ \bibnamefont {Smit}},\
  }\href {\doibase 10.1103/PhysRevLett.76.2613} {\bibfield  {journal} {\bibinfo
   {journal} {Phys. Rev. Lett.}\ }\textbf {\bibinfo {volume} {76}},\ \bibinfo
  {pages} {2613} (\bibinfo {year} {1996})}\BibitemShut {NoStop}%
\bibitem [{\citenamefont {Pawley}\ \emph {et~al.}(1984)\citenamefont {Pawley},
  \citenamefont {Swendsen}, \citenamefont {Wallace},\ and\ \citenamefont
  {Wilson}}]{Pawley_MCRG}%
  \BibitemOpen
  \bibfield  {author} {\bibinfo {author} {\bibfnamefont {G.~S.}\ \bibnamefont
  {Pawley}}, \bibinfo {author} {\bibfnamefont {R.~H.}\ \bibnamefont
  {Swendsen}}, \bibinfo {author} {\bibfnamefont {D.~J.}\ \bibnamefont
  {Wallace}}, \ and\ \bibinfo {author} {\bibfnamefont {K.~G.}\ \bibnamefont
  {Wilson}},\ }\href {\doibase 10.1103/PhysRevB.29.4030} {\bibfield  {journal}
  {\bibinfo  {journal} {Phys. Rev. B}\ }\textbf {\bibinfo {volume} {29}},\
  \bibinfo {pages} {4030} (\bibinfo {year} {1984})}\BibitemShut {NoStop}%
\bibitem [{\citenamefont {Guida}\ and\ \citenamefont
  {Zinn-Justin}(1998)}]{expan1998}%
  \BibitemOpen
  \bibfield  {author} {\bibinfo {author} {\bibfnamefont {R.}~\bibnamefont
  {Guida}}\ and\ \bibinfo {author} {\bibfnamefont {J.}~\bibnamefont
  {Zinn-Justin}},\ }\href@noop {} {\bibfield  {journal} {\bibinfo  {journal}
  {J. Phys. A}\ }\textbf {\bibinfo {volume} {31}},\ \bibinfo {pages} {8103}
  (\bibinfo {year} {1998})}\BibitemShut {NoStop}%
\bibitem [{\citenamefont {Pogorelov}\ and\ \citenamefont
  {Suslov}(2008)}]{expan2008}%
  \BibitemOpen
  \bibfield  {author} {\bibinfo {author} {\bibfnamefont {A.~A.}\ \bibnamefont
  {Pogorelov}}\ and\ \bibinfo {author} {\bibfnamefont {I.~M.}\ \bibnamefont
  {Suslov}},\ }\href@noop {} {\bibfield  {journal} {\bibinfo  {journal} {J.
  Exp. Theor. Phys.}\ }\textbf {\bibinfo {volume} {106}},\ \bibinfo {pages}
  {1118} (\bibinfo {year} {2008})}\BibitemShut {NoStop}%
\bibitem [{\citenamefont {Butera}\ and\ \citenamefont
  {Comi}(2002)}]{HT_expan2002}%
  \BibitemOpen
  \bibfield  {author} {\bibinfo {author} {\bibfnamefont {P.}~\bibnamefont
  {Butera}}\ and\ \bibinfo {author} {\bibfnamefont {M.}~\bibnamefont {Comi}},\
  }\href {\doibase 10.1103/PhysRevB.65.144431} {\bibfield  {journal} {\bibinfo
  {journal} {Phys. Rev. B}\ }\textbf {\bibinfo {volume} {65}},\ \bibinfo
  {pages} {144431} (\bibinfo {year} {2002})}\BibitemShut {NoStop}%
\bibitem [{\citenamefont {Pelissetto}\ and\ \citenamefont
  {Vicari}(2002)}]{CB_review}%
  \BibitemOpen
  \bibfield  {author} {\bibinfo {author} {\bibfnamefont {A.}~\bibnamefont
  {Pelissetto}}\ and\ \bibinfo {author} {\bibfnamefont {E.}~\bibnamefont
  {Vicari}},\ }\href@noop {} {\bibfield  {journal} {\bibinfo  {journal} {Phys.
  Rep.}\ }\textbf {\bibinfo {volume} {368}},\ \bibinfo {pages} {549} (\bibinfo
  {year} {2002})}\BibitemShut {NoStop}%
\bibitem [{\citenamefont {Rosengren}(1986)}]{Rosengren_exact}%
  \BibitemOpen
  \bibfield  {author} {\bibinfo {author} {\bibfnamefont {A.}~\bibnamefont
  {Rosengren}},\ }\href@noop {} {\bibfield  {journal} {\bibinfo  {journal} {J.
  Phys. A: Math. Gen.}\ }\textbf {\bibinfo {volume} {19}},\ \bibinfo {pages}
  {1709} (\bibinfo {year} {1986})}\BibitemShut {NoStop}%
\bibitem [{\citenamefont {Fisher}(1995)}]{Fisher_1995}%
  \BibitemOpen
  \bibfield  {author} {\bibinfo {author} {\bibfnamefont {M.~E.}\ \bibnamefont
  {Fisher}},\ }\href {http://stacks.iop.org/0305-4470/28/i=22/a=009} {\bibfield
   {journal} {\bibinfo  {journal} {J. Phys. A: Math. Gen.}\ }\textbf {\bibinfo
  {volume} {28}},\ \bibinfo {pages} {6323} (\bibinfo {year}
  {1995})}\BibitemShut {NoStop}%
\bibitem [{\citenamefont {El-Showk}\ \emph {et~al.}(2012)\citenamefont
  {El-Showk}, \citenamefont {Paulos}, \citenamefont {Poland}, \citenamefont
  {Rychkov}, \citenamefont {Simmons-Duffin},\ and\ \citenamefont
  {Vichi}}]{conformal_BT2012}%
  \BibitemOpen
  \bibfield  {author} {\bibinfo {author} {\bibfnamefont {S.}~\bibnamefont
  {El-Showk}}, \bibinfo {author} {\bibfnamefont {M.~F.}\ \bibnamefont
  {Paulos}}, \bibinfo {author} {\bibfnamefont {D.}~\bibnamefont {Poland}},
  \bibinfo {author} {\bibfnamefont {S.}~\bibnamefont {Rychkov}}, \bibinfo
  {author} {\bibfnamefont {D.}~\bibnamefont {Simmons-Duffin}}, \ and\ \bibinfo
  {author} {\bibfnamefont {A.}~\bibnamefont {Vichi}},\ }\href@noop {}
  {\bibfield  {journal} {\bibinfo  {journal} {Phys. Rev. D}\ }\textbf {\bibinfo
  {volume} {86}},\ \bibinfo {pages} {025022} (\bibinfo {year}
  {2012})}\BibitemShut {NoStop}%
\bibitem [{\citenamefont {El-Showk}\ \emph {et~al.}(2014)\citenamefont
  {El-Showk}, \citenamefont {Paulos}, \citenamefont {Poland}, \citenamefont
  {Rychkov}, \citenamefont {Simmons-Duffin},\ and\ \citenamefont
  {Vichi}}]{conformal_BT2014}%
  \BibitemOpen
  \bibfield  {author} {\bibinfo {author} {\bibfnamefont {S.}~\bibnamefont
  {El-Showk}}, \bibinfo {author} {\bibfnamefont {M.~F.}\ \bibnamefont
  {Paulos}}, \bibinfo {author} {\bibfnamefont {D.}~\bibnamefont {Poland}},
  \bibinfo {author} {\bibfnamefont {S.}~\bibnamefont {Rychkov}}, \bibinfo
  {author} {\bibfnamefont {D.}~\bibnamefont {Simmons-Duffin}}, \ and\ \bibinfo
  {author} {\bibfnamefont {A.}~\bibnamefont {Vichi}},\ }\href@noop {}
  {\bibfield  {journal} {\bibinfo  {journal} {J. Stat. Phys.}\ }\textbf
  {\bibinfo {volume} {157}},\ \bibinfo {pages} {869} (\bibinfo {year}
  {2014})}\BibitemShut {NoStop}%
\bibitem [{\citenamefont {Kos}\ \emph {et~al.}(2016)\citenamefont {Kos},
  \citenamefont {Poland}, \citenamefont {Simmons-Duffin},\ and\ \citenamefont
  {Vichi}}]{conformal_BT2016}%
  \BibitemOpen
  \bibfield  {author} {\bibinfo {author} {\bibfnamefont {F.}~\bibnamefont
  {Kos}}, \bibinfo {author} {\bibfnamefont {D.}~\bibnamefont {Poland}},
  \bibinfo {author} {\bibfnamefont {D.}~\bibnamefont {Simmons-Duffin}}, \ and\
  \bibinfo {author} {\bibfnamefont {A.}~\bibnamefont {Vichi}},\ }\href@noop {}
  {\bibfield  {journal} {\bibinfo  {journal} {J. High Energ. Phys.}\ }\textbf
  {\bibinfo {volume} {2016}},\ \bibinfo {pages} {36} (\bibinfo {year}
  {2016})}\BibitemShut {NoStop}%
\bibitem [{\citenamefont {Hasenbusch}(2010)}]{Hasenbusch2010}%
  \BibitemOpen
  \bibfield  {author} {\bibinfo {author} {\bibfnamefont {M.}~\bibnamefont
  {Hasenbusch}},\ }\href@noop {} {\bibfield  {journal} {\bibinfo  {journal}
  {Phys. Rev. B}\ }\textbf {\bibinfo {volume} {82}},\ \bibinfo {pages} {174433}
  (\bibinfo {year} {2010})}\BibitemShut {NoStop}%
\bibitem [{\citenamefont {Kaupu{\~z}s}\ \emph {et~al.}(2011)\citenamefont
  {Kaupu{\~z}s}, \citenamefont {Rimsans},\ and\ \citenamefont
  {Melnik}}]{MC_non_per1}%
  \BibitemOpen
  \bibfield  {author} {\bibinfo {author} {\bibfnamefont {J.}~\bibnamefont
  {Kaupu{\~z}s}}, \bibinfo {author} {\bibfnamefont {J.}~\bibnamefont
  {Rimsans}}, \ and\ \bibinfo {author} {\bibfnamefont {R.~V.~N.}\ \bibnamefont
  {Melnik}},\ }\href@noop {} {\bibfield  {journal} {\bibinfo  {journal} {Ukr.
  J. Phys.}\ }\textbf {\bibinfo {volume} {56}},\ \bibinfo {pages} {845}
  (\bibinfo {year} {2011})}\BibitemShut {NoStop}%
\bibitem [{\citenamefont {Kaupu{\~z}s}\ \emph {et~al.}(2014)\citenamefont
  {Kaupu{\~z}s}, \citenamefont {Melnik},\ and\ \citenamefont
  {Rimsans}}]{MC_non_per}%
  \BibitemOpen
  \bibfield  {author} {\bibinfo {author} {\bibfnamefont {J.}~\bibnamefont
  {Kaupu{\~z}s}}, \bibinfo {author} {\bibfnamefont {R.~V.~N.}\ \bibnamefont
  {Melnik}}, \ and\ \bibinfo {author} {\bibfnamefont {J.}~\bibnamefont
  {Rimsans}},\ }\href@noop {} {\bibfield  {journal} {\bibinfo  {journal} {ArXiv
  e-prints}\ } (\bibinfo {year} {2014})},\ \Eprint
  {http://arxiv.org/abs/1407.3095} {arXiv:1407.3095 [cond-mat.stat-mech]}
  \BibitemShut {NoStop}%
\bibitem [{\citenamefont {Wang}\ \emph {et~al.}(2014)\citenamefont {Wang},
  \citenamefont {Xie}, \citenamefont {Chen}, \citenamefont {Normand},\ and\
  \citenamefont {Xiang}}]{tensor_RG_HO}%
  \BibitemOpen
  \bibfield  {author} {\bibinfo {author} {\bibfnamefont {S.}~\bibnamefont
  {Wang}}, \bibinfo {author} {\bibfnamefont {Z.-Y.}\ \bibnamefont {Xie}},
  \bibinfo {author} {\bibfnamefont {J.}~\bibnamefont {Chen}}, \bibinfo {author}
  {\bibfnamefont {B.}~\bibnamefont {Normand}}, \ and\ \bibinfo {author}
  {\bibfnamefont {T.}~\bibnamefont {Xiang}},\ }\href@noop {} {\bibfield
  {journal} {\bibinfo  {journal} {Chin. Phys. Lett.}\ }\textbf {\bibinfo
  {volume} {31}},\ \bibinfo {pages} {070503} (\bibinfo {year}
  {2014})}\BibitemShut {NoStop}%
\bibitem [{\citenamefont {Wu}\ \emph {et~al.}(2008{\natexlab{a}})\citenamefont
  {Wu}, \citenamefont {McCoy}, \citenamefont {Fisher},\ and\ \citenamefont
  {Chayes}}]{Wu_comment}%
  \BibitemOpen
  \bibfield  {author} {\bibinfo {author} {\bibfnamefont {F.}~\bibnamefont
  {Wu}}, \bibinfo {author} {\bibfnamefont {B.~M.}\ \bibnamefont {McCoy}},
  \bibinfo {author} {\bibfnamefont {M.~E.}\ \bibnamefont {Fisher}}, \ and\
  \bibinfo {author} {\bibfnamefont {L.}~\bibnamefont {Chayes}},\ }\href@noop {}
  {\bibfield  {journal} {\bibinfo  {journal} {Phil. Mag.}\ }\textbf {\bibinfo
  {volume} {88}},\ \bibinfo {pages} {3093} (\bibinfo {year}
  {2008}{\natexlab{a}})}\BibitemShut {NoStop}%
\bibitem [{\citenamefont {Wu}\ \emph {et~al.}(2008{\natexlab{b}})\citenamefont
  {Wu}, \citenamefont {McCoy}, \citenamefont {Fisher},\ and\ \citenamefont
  {Chayes}}]{Wu_rejoinder}%
  \BibitemOpen
  \bibfield  {author} {\bibinfo {author} {\bibfnamefont {F.}~\bibnamefont
  {Wu}}, \bibinfo {author} {\bibfnamefont {B.~M.}\ \bibnamefont {McCoy}},
  \bibinfo {author} {\bibfnamefont {M.~E.}\ \bibnamefont {Fisher}}, \ and\
  \bibinfo {author} {\bibfnamefont {L.}~\bibnamefont {Chayes}},\ }\href@noop {}
  {\bibfield  {journal} {\bibinfo  {journal} {Phil. Mag.}\ }\textbf {\bibinfo
  {volume} {88}},\ \bibinfo {pages} {3103} (\bibinfo {year}
  {2008}{\natexlab{b}})}\BibitemShut {NoStop}%
\bibitem [{\citenamefont {Fisher}\ and\ \citenamefont
  {Perk}(2016)}]{Fisher_comment2016}%
  \BibitemOpen
  \bibfield  {author} {\bibinfo {author} {\bibfnamefont {M.~E.}\ \bibnamefont
  {Fisher}}\ and\ \bibinfo {author} {\bibfnamefont {J.~H.~H.}\ \bibnamefont
  {Perk}},\ }\href@noop {} {\bibfield  {journal} {\bibinfo  {journal} {Phys.
  Lett. A}\ }\textbf {\bibinfo {volume} {380}},\ \bibinfo {pages} {1339 }
  (\bibinfo {year} {2016})}\BibitemShut {NoStop}%
\bibitem [{\citenamefont {Ferrenberg}\ and\ \citenamefont
  {Swendsen}(1988)}]{histogram_Ferrenberg1988}%
  \BibitemOpen
  \bibfield  {author} {\bibinfo {author} {\bibfnamefont {A.~M.}\ \bibnamefont
  {Ferrenberg}}\ and\ \bibinfo {author} {\bibfnamefont {R.~H.}\ \bibnamefont
  {Swendsen}},\ }\href@noop {} {\bibfield  {journal} {\bibinfo  {journal}
  {Phys. Rev. Lett.}\ }\textbf {\bibinfo {volume} {61}},\ \bibinfo {pages}
  {2635} (\bibinfo {year} {1988})}\BibitemShut {NoStop}%
\bibitem [{\citenamefont {Ferrenberg}\ and\ \citenamefont
  {Swendsen}(1989)}]{histogram_Ferrenberg1989}%
  \BibitemOpen
  \bibfield  {author} {\bibinfo {author} {\bibfnamefont {A.~M.}\ \bibnamefont
  {Ferrenberg}}\ and\ \bibinfo {author} {\bibfnamefont {R.~H.}\ \bibnamefont
  {Swendsen}},\ }\href@noop {} {\bibfield  {journal} {\bibinfo  {journal}
  {Phys. Rev. Lett.}\ }\textbf {\bibinfo {volume} {63}},\ \bibinfo {pages}
  {1195} (\bibinfo {year} {1989})}\BibitemShut {NoStop}%
\bibitem [{\citenamefont {Weigel}\ and\ \citenamefont
  {Janke}(2009)}]{cross_correlation2009}%
  \BibitemOpen
  \bibfield  {author} {\bibinfo {author} {\bibfnamefont {M.}~\bibnamefont
  {Weigel}}\ and\ \bibinfo {author} {\bibfnamefont {W.}~\bibnamefont {Janke}},\
  }\href@noop {} {\bibfield  {journal} {\bibinfo  {journal} {Phys. Rev. Lett.}\
  }\textbf {\bibinfo {volume} {102}},\ \bibinfo {pages} {100601} (\bibinfo
  {year} {2009})}\BibitemShut {NoStop}%
\bibitem [{\citenamefont {Weigel}\ and\ \citenamefont
  {Janke}(2010)}]{cross_correlation2010}%
  \BibitemOpen
  \bibfield  {author} {\bibinfo {author} {\bibfnamefont {M.}~\bibnamefont
  {Weigel}}\ and\ \bibinfo {author} {\bibfnamefont {W.}~\bibnamefont {Janke}},\
  }\href@noop {} {\bibfield  {journal} {\bibinfo  {journal} {Phys. Rev. E}\
  }\textbf {\bibinfo {volume} {81}},\ \bibinfo {pages} {066701} (\bibinfo
  {year} {2010})}\BibitemShut {NoStop}%
\bibitem [{\citenamefont {Fisher}(1971)}]{FSS_Fisher1971}%
  \BibitemOpen
  \bibfield  {author} {\bibinfo {author} {\bibfnamefont {M.~E.}\ \bibnamefont
  {Fisher}},\ }in\ \href@noop {} {\emph {\bibinfo {booktitle} {Critical
  Phenomena}}},\ \bibinfo {editor} {edited by\ \bibinfo {editor} {\bibfnamefont
  {M.~S.}\ \bibnamefont {Green}}}\ (\bibinfo  {publisher} {Academic Press},\
  \bibinfo {address} {New York},\ \bibinfo {year} {1971})\ pp.\ \bibinfo
  {pages} {1--98}\BibitemShut {NoStop}%
\bibitem [{\citenamefont {Fisher}\ and\ \citenamefont
  {Barber}(1972)}]{FSS_Fisher1972}%
  \BibitemOpen
  \bibfield  {author} {\bibinfo {author} {\bibfnamefont {M.~E.}\ \bibnamefont
  {Fisher}}\ and\ \bibinfo {author} {\bibfnamefont {M.~N.}\ \bibnamefont
  {Barber}},\ }\href@noop {} {\bibfield  {journal} {\bibinfo  {journal} {Phys.
  Rev. Lett}\ }\textbf {\bibinfo {volume} {28}},\ \bibinfo {pages} {1516}
  (\bibinfo {year} {1972})}\BibitemShut {NoStop}%
\bibitem [{\citenamefont {Barber}(1983)}]{FSS_Barber1983}%
  \BibitemOpen
  \bibfield  {author} {\bibinfo {author} {\bibfnamefont {M.~N.}\ \bibnamefont
  {Barber}},\ }in\ \href@noop {} {\emph {\bibinfo {booktitle} {Phase
  Transitions and Critical Phenomena}}},\ Vol.~\bibinfo {volume} {8},\ \bibinfo
  {editor} {edited by\ \bibinfo {editor} {\bibfnamefont {C.}~\bibnamefont
  {Domb}}\ and\ \bibinfo {editor} {\bibfnamefont {J.~L.}\ \bibnamefont
  {Lebowitz}}}\ (\bibinfo  {publisher} {Academic Press},\ \bibinfo {address}
  {New York},\ \bibinfo {year} {1983})\ pp.\ \bibinfo {pages}
  {146--266}\BibitemShut {NoStop}%
\bibitem [{\citenamefont {Privman(editor)}(1990)}]{FSS_Privman}%
  \BibitemOpen
  \bibfield  {author} {\bibinfo {author} {\bibfnamefont {V.}~\bibnamefont
  {Privman(editor)}},\ }\href@noop {} {\emph {\bibinfo {title} {Finite-Size
  Scaling and Numerical Simulation}}}\ (\bibinfo  {publisher} {World
  Scientific},\ \bibinfo {address} {Singapore},\ \bibinfo {year}
  {1990})\BibitemShut {NoStop}%
\bibitem [{\citenamefont {Wolff}(1989)}]{Wolff}%
  \BibitemOpen
  \bibfield  {author} {\bibinfo {author} {\bibfnamefont {U.}~\bibnamefont
  {Wolff}},\ }\href {\doibase 10.1103/PhysRevLett.62.361} {\bibfield  {journal}
  {\bibinfo  {journal} {Phys. Rev. Lett.}\ }\textbf {\bibinfo {volume} {62}},\
  \bibinfo {pages} {361} (\bibinfo {year} {1989})}\BibitemShut {NoStop}%
\bibitem [{\citenamefont {Matsumoto}\ and\ \citenamefont
  {Nishimura}(1998)}]{Mersenne_Twister}%
  \BibitemOpen
  \bibfield  {author} {\bibinfo {author} {\bibfnamefont {M.}~\bibnamefont
  {Matsumoto}}\ and\ \bibinfo {author} {\bibfnamefont {T.}~\bibnamefont
  {Nishimura}},\ }\href {\doibase 10.1145/272991.272995} {\bibfield  {journal}
  {\bibinfo  {journal} {ACM Trans. Model. Comput. Simul.}\ }\textbf {\bibinfo
  {volume} {8}},\ \bibinfo {pages} {3} (\bibinfo {year} {1998})}\BibitemShut
  {NoStop}%
\bibitem [{\citenamefont {Ito}\ and\ \citenamefont
  {Kohring}(1993)}]{Ito_Kohring_1993}%
  \BibitemOpen
  \bibfield  {author} {\bibinfo {author} {\bibfnamefont {N.}~\bibnamefont
  {Ito}}\ and\ \bibinfo {author} {\bibfnamefont {G.~A.}\ \bibnamefont
  {Kohring}},\ }\href {\doibase https://doi.org/10.1016/0378-4371(93)90127-P}
  {\bibfield  {journal} {\bibinfo  {journal} {Physica A: Statistical Mechanics
  and its Applications}\ }\textbf {\bibinfo {volume} {201}},\ \bibinfo {pages}
  {547 } (\bibinfo {year} {1993})}\BibitemShut {NoStop}%
\bibitem [{\citenamefont {Kiefer}(1953)}]{golden_section}%
  \BibitemOpen
  \bibfield  {author} {\bibinfo {author} {\bibfnamefont {J.}~\bibnamefont
  {Kiefer}},\ }\href {http://www.jstor.org/stable/2032161} {\bibfield
  {journal} {\bibinfo  {journal} {Proceedings of the American Mathematical
  Society}\ }\textbf {\bibinfo {volume} {4}},\ \bibinfo {pages} {502} (\bibinfo
  {year} {1953})}\BibitemShut {NoStop}%
\bibitem [{\citenamefont {Binder}(1981)}]{cumulant_Binder1981}%
  \BibitemOpen
  \bibfield  {author} {\bibinfo {author} {\bibfnamefont {K.}~\bibnamefont
  {Binder}},\ }\href@noop {} {\bibfield  {journal} {\bibinfo  {journal} {Z.
  Phys. B}\ }\textbf {\bibinfo {volume} {43}},\ \bibinfo {pages} {119}
  (\bibinfo {year} {1981})}\BibitemShut {NoStop}%
\bibitem [{\citenamefont {Aharony}\ and\ \citenamefont
  {Fisher}(1983)}]{Fisher_correction}%
  \BibitemOpen
  \bibfield  {author} {\bibinfo {author} {\bibfnamefont {A.}~\bibnamefont
  {Aharony}}\ and\ \bibinfo {author} {\bibfnamefont {M.~E.}\ \bibnamefont
  {Fisher}},\ }\href {\doibase 10.1103/PhysRevB.27.4394} {\bibfield  {journal}
  {\bibinfo  {journal} {Phys. Rev. B}\ }\textbf {\bibinfo {volume} {27}},\
  \bibinfo {pages} {4394} (\bibinfo {year} {1983})}\BibitemShut {NoStop}%
\bibitem [{\citenamefont {Campostrini}\ \emph {et~al.}(2002)\citenamefont
  {Campostrini}, \citenamefont {Pelissetto}, \citenamefont {Rossi},\ and\
  \citenamefont {Vicari}}]{Massimo_wnr}%
  \BibitemOpen
  \bibfield  {author} {\bibinfo {author} {\bibfnamefont {M.}~\bibnamefont
  {Campostrini}}, \bibinfo {author} {\bibfnamefont {A.}~\bibnamefont
  {Pelissetto}}, \bibinfo {author} {\bibfnamefont {P.}~\bibnamefont {Rossi}}, \
  and\ \bibinfo {author} {\bibfnamefont {E.}~\bibnamefont {Vicari}},\ }\href
  {\doibase 10.1103/PhysRevE.65.066127} {\bibfield  {journal} {\bibinfo
  {journal} {Phys. Rev. E}\ }\textbf {\bibinfo {volume} {65}},\ \bibinfo
  {pages} {066127} (\bibinfo {year} {2002})}\BibitemShut {NoStop}%
\bibitem [{\citenamefont {Efron}\ and\ \citenamefont
  {Tibshirani}(1993)}]{jackknife}%
  \BibitemOpen
  \bibfield  {author} {\bibinfo {author} {\bibfnamefont {B.}~\bibnamefont
  {Efron}}\ and\ \bibinfo {author} {\bibfnamefont {R.~J.}\ \bibnamefont
  {Tibshirani}},\ }\href@noop {} {\emph {\bibinfo {title} {An Introduction to
  the Bootstrap}}}\ (\bibinfo  {publisher} {Chapman and Hall},\ \bibinfo
  {address} {New York},\ \bibinfo {year} {1993})\BibitemShut {NoStop}%
\bibitem [{\citenamefont {Guskova}\ \emph {et~al.}(2016)\citenamefont
  {Guskova}, \citenamefont {Barash},\ and\ \citenamefont {Shchur}}]{Guskova}%
  \BibitemOpen
  \bibfield  {author} {\bibinfo {author} {\bibfnamefont {M.~S.}\ \bibnamefont
  {Guskova}}, \bibinfo {author} {\bibfnamefont {L.~Y.}\ \bibnamefont {Barash}},
  \ and\ \bibinfo {author} {\bibfnamefont {L.~N.}\ \bibnamefont {Shchur}},\
  }\href@noop {} {\bibfield  {journal} {\bibinfo  {journal} {Computer Physics
  Communications}\ }\textbf {\bibinfo {volume} {200}},\ \bibinfo {pages} {402}
  (\bibinfo {year} {2016})}\BibitemShut {NoStop}%
\bibitem [{\citenamefont {Ferrenberg}\ \emph {et~al.}(1992)\citenamefont
  {Ferrenberg}, \citenamefont {Landau},\ and\ \citenamefont {Wong}}]{FLWrng}%
  \BibitemOpen
  \bibfield  {author} {\bibinfo {author} {\bibfnamefont {A.}~\bibnamefont
  {Ferrenberg}}, \bibinfo {author} {\bibfnamefont {D.~P.}\ \bibnamefont
  {Landau}}, \ and\ \bibinfo {author} {\bibfnamefont {Y.}~\bibnamefont
  {Wong}},\ }\href@noop {} {\bibfield  {journal} {\bibinfo  {journal} {Phys.
  Rev. Lett.}\ }\textbf {\bibinfo {volume} {69}},\ \bibinfo {pages} {3382}
  (\bibinfo {year} {1992})}\BibitemShut {NoStop}%
\bibitem [{\citenamefont {Wang}\ \emph {et~al.}(2013)\citenamefont {Wang},
  \citenamefont {Zhou}, \citenamefont {Zhang}, \citenamefont {Garoni},\ and\
  \citenamefont {Deng}}]{deng_percolation}%
  \BibitemOpen
  \bibfield  {author} {\bibinfo {author} {\bibfnamefont {J.}~\bibnamefont
  {Wang}}, \bibinfo {author} {\bibfnamefont {Z.}~\bibnamefont {Zhou}}, \bibinfo
  {author} {\bibfnamefont {W.}~\bibnamefont {Zhang}}, \bibinfo {author}
  {\bibfnamefont {T.~M.}\ \bibnamefont {Garoni}}, \ and\ \bibinfo {author}
  {\bibfnamefont {Y.}~\bibnamefont {Deng}},\ }\href {\doibase
  10.1103/PhysRevE.87.052107} {\bibfield  {journal} {\bibinfo  {journal} {Phys.
  Rev. E}\ }\textbf {\bibinfo {volume} {87}},\ \bibinfo {pages} {052107}
  (\bibinfo {year} {2013})}\BibitemShut {NoStop}%
\bibitem [{\citenamefont {Bl\"ote}\ \emph {et~al.}(1999)\citenamefont
  {Bl\"ote}, \citenamefont {Shchur},\ and\ \citenamefont
  {Talapov}}]{cluster_processor}%
  \BibitemOpen
  \bibfield  {author} {\bibinfo {author} {\bibfnamefont {H.~W.~J.}\
  \bibnamefont {Bl\"ote}}, \bibinfo {author} {\bibfnamefont {L.~N.}\
  \bibnamefont {Shchur}}, \ and\ \bibinfo {author} {\bibfnamefont {A.~L.}\
  \bibnamefont {Talapov}},\ }\href@noop {} {\bibfield  {journal} {\bibinfo
  {journal} {Int. J. Mod. Phys. C}\ }\textbf {\bibinfo {volume} {10}},\
  \bibinfo {pages} {1137} (\bibinfo {year} {1999})}\BibitemShut {NoStop}%
\bibitem [{\citenamefont {Deng}\ and\ \citenamefont
  {Bl\"ote}(2003)}]{Blote_Deng_2003}%
  \BibitemOpen
  \bibfield  {author} {\bibinfo {author} {\bibfnamefont {Y.}~\bibnamefont
  {Deng}}\ and\ \bibinfo {author} {\bibfnamefont {H.~W.~J.}\ \bibnamefont
  {Bl\"ote}},\ }\href {\doibase 10.1103/PhysRevE.68.036125} {\bibfield
  {journal} {\bibinfo  {journal} {Phys. Rev. E}\ }\textbf {\bibinfo {volume}
  {68}},\ \bibinfo {pages} {036125} (\bibinfo {year} {2003})}\BibitemShut
  {NoStop}%
\bibitem [{\citenamefont {Gaunt}(1982)}]{Gaunt_1982}%
  \BibitemOpen
  \bibfield  {author} {\bibinfo {author} {\bibfnamefont {D.~S.}\ \bibnamefont
  {Gaunt}},\ }in\ \href@noop {} {\emph {\bibinfo {booktitle} {Phase
  Transitions, Proc. 1980 Cargese Summer Institute}}},\ \bibinfo {editor}
  {edited by\ \bibinfo {editor} {\bibfnamefont {M.}~\bibnamefont {Levy}},
  \bibinfo {editor} {\bibfnamefont {J.~C.}\ \bibnamefont {{Le Guillou}}}, \
  and\ \bibinfo {editor} {\bibfnamefont {J.}~\bibnamefont {Zinn-Justin}}}\
  (\bibinfo  {publisher} {Plenum},\ \bibinfo {address} {New York},\ \bibinfo
  {year} {1982})\BibitemShut {NoStop}%
\bibitem [{\citenamefont {Adler}(1983)}]{Adler_1983}%
  \BibitemOpen
  \bibfield  {author} {\bibinfo {author} {\bibfnamefont {J.}~\bibnamefont
  {Adler}},\ }\href@noop {} {\bibfield  {journal} {\bibinfo  {journal} {J.
  Phys. A : Math. Gen.}\ }\textbf {\bibinfo {volume} {16}},\ \bibinfo {pages}
  {3585} (\bibinfo {year} {1983})}\BibitemShut {NoStop}%
\end{thebibliography}%

\end{document}